\documentclass[aps,prb,titlepage,twocolumn,showpacs,amsmath,amssymb,floatfix,superscriptaddress]{revtex4}

\usepackage{graphicx}
\usepackage{amsmath,amssymb}
\usepackage[T1]{fontenc}
\usepackage[latin1]{inputenc}
\usepackage{times}
\usepackage{bm}
\usepackage[usenames,dvips]{color}
\usepackage[normalem]{ulem}
\usepackage{setspace}
  \makeatletter
  \def\@dotsep{4.5}
  \makeatother
\newlength{\myVSpace}
\newcommand{\be}{\begin{equation}}
\newcommand{\ee}{\end{equation}}
\newcommand{\ba}{\begin{eqnarray}}
\newcommand{\ea}{\end{eqnarray}}

\begin{document}

\title{Towards optimized suppression of dephasing in systems subject to pulse timing constraints}

\author{Thomas E. Hodgson}
\affiliation{Department of Physics, University of York, Heslington, York,
YO10 5DD, United Kingdom}
\author{Lorenza Viola}
\affiliation{\mbox{Department of Physics and Astronomy, 
6127 Wilder Laboratory, Dartmouth College, Hanover, New Hampshire 03755, USA}}
\author{Irene D'Amico}
\affiliation{Department of Physics, University of York, Heslington, York,
YO10 5DD, United Kingdom}

\begin{abstract}
We investigate the effectiveness of different dynamical decoupling
protocols for storage of a single qubit in the presence of a purely
dephasing bosonic bath, with emphasis on comparing quantum coherence
preservation under uniform vs. non-uniform delay times between pulses.
In the limit of instantaneous bit-flip pulses, this is accomplished by
establishing a new representation of the controlled qubit evolution,
where the resulting decoherence behaviour is directly expressed in
terms of the free evolution.  Simple analytical expressions are given
to approximate the long- and short- term coherence behaviour for both
ohmic and supra-ohmic environments. We focus on systems with physical
constraints on achievable time delays, with emphasis on pure dephasing
of excitonic qubits in quantum dots.  Our analysis shows that 
little advantage of high-level decoupling schemes based on
concatenated or optimal design is to be expected if operational
constraints prevent pulses to be applied sufficiently fast.  In such
constrained scenarios, we demonstrate how simple modifications of
repeated periodic echo protocols can offer significantly improved
coherence preservation in realistic parameter regimes.
\end{abstract}

\date{\today}


\maketitle


\section{Introduction}

The ability to effectively counteract decoherence processes in
physical quantum information processing (QIP) devices is a fundamental
prerequisite for taking advantage of the added power promised by
quantum computation and quantum simulation as compared to purely
classical methods.  Dynamical decoupling (DD) techniques for open
quantum systems \cite{lorenza,Viola:1999:2417} have been shown to be
able to significantly suppress non-Markovian decoherence for storage
times that can be very long relative to the typical time scales
associated with the decoherence process itself.  Over the last decade,
the design and characterization of viable DD schemes for realistic
qubit devices has spurred an intense theoretical and experimental
effort, taking DD well beyond the original nuclear magnetic resonance
(NMR) setting \cite{NMR}. While earlier DD schemes relied on the
simple periodic repetition of instantaneous pulses (so-called
`bang-bang' periodic DD, PDD \cite{Viola:1999:2417}, and its
closely-related time-symmetrized version, so-called Carr-Purcell DD,
CPDD \cite{carr-purcell,NMR}), recent theoretical investigations have
explored the benefits of more sophisticated control design in a number
of ways.  In particular, this has led to devising recursive and
randomized pulse sequences for generic decoherence models on
finite-dimensional quantum systems -- so-called `concatenated' DD (CDD
\cite{khodjastehprl,khodjasteh}) and `randomized' DD
\cite{Viola2005Random,Lea}; to identifying `optimal' protocols for a
single qubit undergoing pure dephasing -- most notably, the
so-called Uhrig DD (UDD) \cite{Dhar,opt,opt-njp,opt-spc,lee08,yang},
and its extension to `locally optimized' \cite{nist1,nist2} DD
sequences tailored to specific noise environments; and, most recently,
to combining the advantages of concatenation and optimization for a
single qubit exposed to arbitrary decoherence
\cite{cudd,qdd,gerschon}.  As a key common feature, these
investigations highlight the sensitivity of DD performance to the
details of the applied control path, and point to the importance of
carefully tuning the relative pulse delays in order to boost the
efficiency of the achievable decoherence suppression \cite{remark1}.

In view of the above rich scenario, assessing the performance of
different DD protocols {\em in specific qubit devices and/or in the
presence of specific control constraints} becomes especially
important. Recently, the effectiveness of traditional multi-pulse
spin-echo sequences based on PDD and CPDD, as compared to `high-level'
protocols based on CDD and UDD, has been scrutinized in several
control settings. In particular, a number of theoretical studies have
addressed suppression of pure dephasing associated to spectral
diffusion \cite{witz08} and hyperfine-induced decoherence \cite{wen}
from a quantum  spin-bath for an electron-spin qubit, as well
as suppression of classical $1/f$ phase noise in a superconducting
qubit \cite{lara,nave}.  Experimentally, the performance of CDD
protocols has been characterized for an NMR spin qubit \cite{Suter},
while optimal UDD implementations have been reported for both a
trapped ion qubit exposed to engineered classical phase noise
\cite{nist1,nist2,nist3} and, most recently, for electron spin qubits
undergoing spin-bath decoherence in a malonic acid crystal
\cite{Liu09}.  These studies have demonstrated, in particular, how UDD
can significantly outperform low-level DD schemes provided that the
noise spectrum has a sharp high-frequency cutoff and sufficiently high
pulse repetition rates may be afforded.  

Amongst prospective solid-state QIP platforms, exciton qubits in
self-assembled quantum dots (QDs) have likewise received vast
attention in recent years \cite{irene,exciton}: due to the coupling to
photons, excitons can be driven all-optically on sub-picosecond time
scales\cite{irene}. Excitonic implementations also allow the flexibility of
designing hybrid solid state-flying qubit schemes
\cite{MangQED,Imamoglu}.  Pure dephasing turns out to be the dominant
factor limiting the coherence lifetime in such qubit devices, where
strong coupling with phonon modes of the host crystal result in
typical decoherence ($T_2$) time scales of a few pico-seconds
\cite{1stkuhn}.  We have previously shown in
Ref. \onlinecite{dephasing} that, remarkably, PDD allows for
substantial exciton coherence recovery in experimentally relevant
parameter regimes (up to $90\%$ recovery over $\sim 10$ ps at room
temperature), the control performance being especially enhanced for QD
shapes and bias fields optimized for quantum computing architectures.
Our goal in this paper is to quantitatively assess to what extent more
elaborated DD schemes -- in particular, sequences employing
non-uniform pulse timings -- can improve beyond the simplest PDD
setting when a {\em lower bound on the achievable control time scale}
(minimum pulse separation) is present. 
 
We find that in the presence of such a timing limitation, simple
protocols such as PDD or CPDD may outperform high-level sequences
based on CDD/UDD.  Interestingly, on the one hand this reinforces
similar conclusions drawn in Ref. \onlinecite{nave} for classical
dephasing in superconducting qubits.  On the other hand, we
additionally show how it is possible to engineer a suitable
`preparatory' sequence  that enhances the performance of a
subsequent PDD pulse train.  In the process, we take advantage of the
exact solvability of a purely dephasing model in the presence of
instantaneous pulses to obtain an exact representation of the
controlled dynamics in terms of the free evolution.  This allows
rigorous results on the long-time asymptotic decoherence behaviour to
be established for generic noise spectral densities, by allowing in
particular a comparison between ohmic and supra-ohmic environments.
Furthermore, our work provides a first explicit analysis of CDD
performance in the presence of a quantum bosonic bath.  From a
practical standpoint, our results suggest that simple DD protocols may
remain a method of choice if significant timing constraints are in
place, and that incorporating such constraints from the outset is
necessary before further optimization can show its benefits.  While
our numerical results are tailored to excitons in QDs, we expect the
above conclusions to be relevant for other constrained qubit devices.

\section{Single-qubit dephasing dynamics}
\label{basic}

We consider the pure dephasing dynamics of a single qubit coupled to a
non-interacting bath of harmonic oscillators.  The Hamiltonian of such
a system may be written in the form 
\begin{eqnarray} 
H &=&\frac{E}{2} \sigma_z +\hbar\sum_{j} \omega_j
b^{\dagger}_{j}b_{j}\nonumber \\ &+& \hbar\sum_{j}
(g^*_{j}b^{\dagger}_{j}+g_{j}b_{j})[(1-\alpha)\sigma_0+\alpha\sigma_z]
\label{arbhamiltonian}\\ &\equiv & H_0+\hbar\sum_{j}
(g^*_{j}b^{\dagger}_{j}+g_{j}b_{j})[(1-\alpha)\sigma_0+\alpha\sigma_z],
\ea
\noindent 
where $E$ gives the energy difference between the qubit's levels,
$b^{\dagger}_j$ and $b_j$ are canonical creation and annihilation
operators of the oscillator mode $j$, 
and $g_j$ describes the coupling between the qubit and the $j$-th bath
mode.  In the above expression for $H$, the parameter $\alpha$
accounts for the possibility that either both or only one of the spin
(or pseudo-spin) qubit computational levels effectively couple to the
bath: $\alpha=1$ corresponds to the standard purely-dephasing
spin-boson model, whereas if $\alpha=1/2$, only the $\sigma_z=+1$
eigenstate couples to the bath.  Specifically, for an excitonic qubit,
the logical states are represented by the presence or absence a single
(ground-state) exciton in the QD \cite{irene}, and $E$ is
the energy relative to the crystal ground state.

As time evolves, the qubit becomes entangled with the environment and
the off-diagonal elements of the qubit density matrix evaluated at
time $t$ in the interaction picture with respect to $H_0$ read
\cite{lorenza,1stkuhn}
\begin{equation}
\rho_{01}(t) =\rho_{10}^* (t) =\rho_{01}(t=0)e^{-\Gamma (t)},
\label{dens}
\end{equation}
\begin{eqnarray}
\Gamma(t)&\equiv &\Gamma_0(t)\label{Gam1}\\
&=&(2\alpha)^2\int_{0}^{\infty}d\omega\frac{
I(\omega)}{\omega^2}\coth\Big(\frac{\hbar\omega}{2k_B
T}\Big)[1-\cos(\omega t)],\nonumber
\end{eqnarray}
where $T$ is the temperature, $k_B$ the Boltzmann's constant, and
\begin{equation}
I(\omega)= \sum_{j}\delta(\omega-\omega_j)|g_{j}|^2 
\label{Iom}
\end{equation}
is the spectral density function characterizing the interaction of the
qubit with the oscillator bath. For a supra-ohmic environment,
$I(\omega)\stackrel{\omega\to0}{\sim}\omega^3$, as opposed, for
instance, to an ohmic reservoir where
$I(\omega)\stackrel{\omega\to0}{\sim}\omega$.  Likewise, the
high-frequency behaviour is characterized by a frequency cut-off
$\omega_c$, for instance, for excitons one can assume that
$I(\omega)\stackrel{\omega\to \infty}{\sim}e^{-\omega^2/\omega_c^2}$.

As it turns out, the decoherence of the qubit in the presence of an
{\em arbitrary} sequence of bang-bang pulses, each effecting an
instantaneous $\pi$ rotation, can still be exactly described by
Eq.~(\ref{dens}), provided a modified decoherence function is used
\cite{Viola:1999:2417,Uchiyama:2002:032313,opt}. Consider an arbitrary
storage time $t$, during which a total number $s$ of pulses is
applied, at instants $\{t_1,\ldots,t_n,\ldots,t_s\}$, with $0< t_1 <
t_2 <\ldots t_s <t$.  By using the theory developed by Uhrig in
Refs.~\onlinecite{opt,opt-njp}, we can define a controlled coherence
function $\Gamma(t)$ in the following way:
\begin{eqnarray}
\label{gammaopt}
\Gamma(t) \equiv 
\left\{ \begin{array}{ll}
\Gamma_0 (t) & \;\; t \le t_1, \\
\Gamma_n (t) & \;\; t_n < t \le t_{n+1},\; 0<n<s, \\
\Gamma_s (t) & \;\; t_s <t .
\end{array} \right.
\end{eqnarray}  
Here, $\Gamma_0(t)$ is given in Eq. (\ref{Gam1}) whereas for
$1\leq n\leq s$ we let \cite{opt}
\begin{eqnarray*}
\Gamma_n (t) & = &(2\alpha)^2\int^{\infty}_0
\frac{I(\omega)}{2\omega^2}\coth\Big(\frac{\hbar
\omega}{2k_BT}\Big)|y_n(\omega t)|^2d \omega, \\
y_n (z) & = &1+(-1)^{n+1}e^{iz}+2\sum^n_{m=1}(-1)^me^{iz\delta_m},\;\;
z >0,
\end{eqnarray*} 
with the $n$-th pulse being understood to occur at time $t_n=\delta_n
t$, and $0<\delta_1<\ldots \delta_n< \ldots \delta_s<1$.  While the
instantaneous pulse assumption must be handled with care in general,
we have discussed in Ref.~\onlinecite{dephasing} how it translates
into reasonable physical constraints for an excitonic qubit coupled to
a phononic bath.

We now proceed to directly relate $\Gamma_n(t)$ to $\Gamma_0(t)$ for
arbitrary $n$. Let us first rewrite the above coherence function
$\Gamma_n(t)$ in a compact way as
\begin{equation}
\Gamma_n(t)=\int_0^\infty\eta(\omega) |y_n(\omega t)|^2d\omega, \;\;\;
n\ge 0,
\label{gammaneta}
\end{equation}
where we have defined 
\be |y_0(\omega t)|^2\equiv |1- e^{i\omega t}|^2, 
\ee 
and
\begin{equation}
\eta(\omega)=(2\alpha)^2\frac{I(\omega)}{2\omega^2}\coth\Big(\frac{\hbar
\omega}{2k_BT} \Big).
\label{eta}\end{equation}
By relating $|y_1(\omega t)|^2$ to $|y_0(\omega t)|^2$ we can write
\begin{equation}
\Gamma_1(t)=-\Gamma_0(t)+2\Gamma_0(t_1)+2\Gamma_0(t-t_1).\label{gamma10}
\end{equation}
Upon continuing this iteration we find
\begin{eqnarray}
\Gamma_2(t)&\hspace*{-0.8mm}=\hspace*{-0.8mm}&-\Gamma_1(t)+2\Gamma_1(t_2)+2
\Gamma_0(t-t_1)\nonumber,
\\ \mbox{} & \hspace*{-0.8mm}\ldots \hspace*{-0.8mm}& \mbox{}\nonumber
\\\label{Gn-1}
\Gamma_n(t)&\hspace*{-0.8mm}=\hspace*{-0.8mm}&-\Gamma_{n-1}(t)+2
\Gamma_{n-1}(t_{n})+2\Gamma_0(t-t_n). 
\end{eqnarray}
Furthermore, by expressing $|y_n(\omega t)|^2$ as a function of
$|y_0(\omega t)|^2$, we are able to write the entire evolution in the
presence of an {\it arbitrary} pulse sequence only in terms of the
uncontrolled evolution.  Explicitly, we find:
\begin{eqnarray}
\hspace*{-2mm}\Gamma_n(t)&\hspace*{-1.5mm}=\hspace*{-1.5mm}&
2\sum_{m=1}^n(-1)^{m+1}\Gamma_0(t_m) \nonumber\\
\hspace*{-1.5mm}&+\hspace*{-1.5mm}&
4\sum_{m=2}^{n}\sum_{j<m}\Gamma_0(t_m-t_j)(-1)^{m-1+j}\nonumber\\
&\hspace*{-1.5mm}+\hspace*{-1.5mm}&2\sum_{m=1}^n(-1)^{m+n}\Gamma_0(t-t_m)
+(-1)^n\Gamma_0(t).
\label{Gamman-1}
\end{eqnarray}
The above equation is one of the main results of this paper.  By using
Eq.~(\ref{Gamman-1}), it is, in particular, straightforward to see
that \be \Gamma_{n-1}(t_n)=\lim_{t\to
t_n}\Gamma_{n}(t)=\Gamma_{n}(t_n).
\label{n_eq_nminus1} \ee 
This confirms that the function $\Gamma(t)$ as defined in
Eqs.~(\ref{gammaopt}) is continuos at the (instantaneous) pulse
timings, as expected on physical grounds.

As a first example of the usefulness of this representation, we
consider how two pulses may be used to increase the asymptotic
coherence of a supra-ohmic system, in which the free dephasing
dynamics saturates in the long-time limit to a {\em finite} value
\cite{palmaproc,1stkuhn} $\Gamma_0(\infty) >0$. Taking the
$t\to\infty$ limit in Eq.~(\ref{gamma10}) or, equivalently, letting
$n=1$ in Eq. (\ref{Gamman-1}), yields
$\Gamma_1(\infty)=2\Gamma_0(t_1)+\Gamma_0(\infty)$.  Since
$\Gamma_0(t) \geq 0$ for all $t$, this shows how a single pulse cannot
decrease the asymptotic decoherence level.  However, after two pulses
we have
\begin{eqnarray}
\Gamma_2(t)&=&\Gamma_0(t)-2\Gamma_0(t-t_1)-2\Gamma_0(t_2)\nonumber\\
&+&2\Gamma_0(t_1)+4\Gamma_0(t_2-t_1)+2\Gamma_0(t-t_2).
\end{eqnarray}
Therefore, 
\begin{equation}
\Gamma_2(\infty)=\Gamma_0(\infty)-2\Gamma_0(t_2)+2\Gamma_0
(t_1)+4\Gamma_0(t_2-t_1),
\end{equation} 
and $t_1$ and $t_2$ can be chosen to decrease the asymptotic
decoherence provided that
\begin{equation}
2\Gamma_0(t_2)-2\Gamma_0(t_1)>4\Gamma_0(t_2-t_1).
\label{condition}
\end{equation}
In Fig. \ref{asymptdec}, we plot $\Gamma(t)$ for an exciton qubit
coupled to phonon modes and subject to two control pulses at
$t_1=0.2$~ps and $t_2=0.31$~ps.  For comparison, we also plot the
evolution under a single control pulse at $t_1=0.2$~ps and the free
evolution $\Gamma_0(t)$. As one can see, Eq.~(\ref{condition}) can
indeed be satisfied. Numerical results showing how a few pulses can
increase the asymptotic coherence have been reported for excitonic
dephasing  in Ref. \onlinecite{kuhncontrol}.

\begin{figure}[t]
\includegraphics*[width=\linewidth]{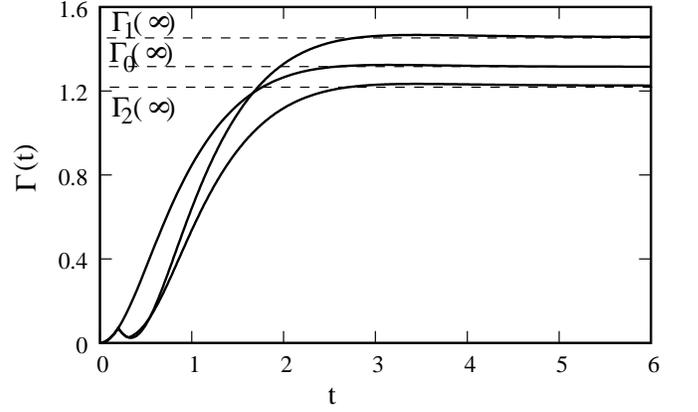}
\caption{Comparison between $\Gamma_0(t)$, $\Gamma_1(t)$, and
$\Gamma_2(t)$ for an exciton qubit at $T=77$ K,  as computed from
Eq. (\ref{gammaopt}).  Pulse times are $t_1=0.2$~ps and $t_2=0.31$~ps. }
\label{asymptdec}
\end{figure}

For the case of Fig.~\ref{asymptdec}, as well as for all the numerical
examples in this paper, we consider (unless otherwise stated) an
exciton qubit tightly confined within a GaAs QD at 77 K. The QD
potentials are modeled as parabolic in all three dimensions, with
confinement energies in the $z$-direction of $\hbar\omega_e=505$~meV
and $\hbar\omega_h=100$~meV, while $\hbar\omega_e=30$~meV and
$\hbar\omega_h=24$~meV in the in-plane
directions\cite{irene,kuhn-irene}. The subscript $e/h$ indicates
electron/hole, respectively. For this exciton, in the absence of
control most of the coherence is lost after a few picoseconds
\cite{kuhn-irene}.  Having this specific system in mind, we shall set
$\alpha=1/2$ henceforth in our numerical calculations, and plot the
quantity $|\exp(-\Gamma_n(t))|^2$, which is directly proportional to
the square modulus of the measured optical polarization ${\bf P}(t)$.

As discussed in detail in Ref. \onlinecite{ dephasing}, the spectral
density of this system is given by
\begin{eqnarray}
I(\omega)=I_e(\omega)+I_h(\omega)+I_{eh}(\omega),
\end{eqnarray}
where the indices $e/h/eh$ correspond to single particle spectral
densities of the electron and the hole, and to the electron-hole
inteference term respectively, and
\begin{eqnarray}
I_{e/h/eh}(\omega)=\sum_i
F_i^{e/h/eh}(\omega)\exp\bigg(-\frac{\omega^2}{\omega_{c_i,e/h/eh}^2}\bigg).
\end{eqnarray}
Here, $i$ labels different phonon modes, whereas
$F_i^{e/h/eh}(\omega)$ is a mode-dependent function for which
$F_i^{e/h/eh}(\omega)\stackrel{\omega\to0}{\leq}\omega^3$.
The spectral density may be further approximated as
\begin{equation}
I(\omega)\approx
F\omega^3\exp\Big(-\frac{\omega^2}{\omega_c^2}\Big),
\label{form}
\end{equation}
where the parameters $F$ and $\omega_c$ are determined by fitting a
curve of the form Eq.~(\ref{form}) to the actual exciton spectral
density. For the particular exciton parameters listed above, this
yields $F=1.14\times10^{-26}s$ and $\hbar\omega_c=2$meV.

\section{Periodic DD: Performance and exact asymptotic properties}
\label{pdd}

For a Hamiltonian as in Eq. (\ref{arbhamiltonian}), a
DD cycle consisting of two uniformly spaced rotations by $\pi$ about
the $x$ axis,
\begin{equation}
X \Delta t X \Delta t,
\label{cycle}
\end{equation}
where time ordering is understood from right to left, removes the
interaction between the qubit and the boson bath
\cite{Viola:1999:2417,khodjasteh} to the lowest (perturbative) order
in $\omega_c T_c$, with $T_c =2\Delta t$.  The simplest DD protocol,
PDD, is obtained by iterating the above control cycle in time.

Fig.~\ref{equidist} compares the free evolution with the
PDD-controlled dephasing for the exciton qubit under examination,
computed from the exact expressions given in Sec \ref{basic}.
Sequences with three different pulse delays are shown, $\Delta
t=0.1$~ps, $\Delta t=0.2$~ps, and $\Delta t=0.3$~ps, respectively. For
the exciton qubit, two conditions determine a suitable range of
$\Delta t$ for effective PDD: i) On the one hand, it is necessary that
the control time scale $T_c$ be sufficiently short with respect to the
(shortest) correlation time of the decoherence dynamics, which means
in this case $2\Delta t \lesssim \tau_c =2\pi/\omega_c$.  Physically,
this can also be interpreted by requiring that the characteristic
frequency introduced by the periodic control,
$$ \omega_{\text{res}}=\frac{\pi}{\Delta t},$$ 
\noindent 
be significantly higher than the spectral cut-off fequency itself,
$\omega_{\text{res}} \gtrsim \omega_c$, in such a way that the
DD-renormalized spectral density function, $I(\omega)\tan^2(\omega
\Delta t/2)$, is effectively `up-shifted' beyond the bath cutoff
\cite{wenJMO,dephasing,opt-spc,gerschon}.  ii) On the other hand, the
existence of a lower bound on the pulse duration implies a lower bound
on the separation $\Delta t$ in order for the instantaneous-pulse
description to be accurate.  As discussed in
Ref. \onlinecite{dephasing}, this means $\Delta t\gtrsim 0.1$~ps for
semiconductor self-assembled QDs of interest for QIP.

The values of $\Delta t$ used in Fig.~\ref{equidist}, are consistent
with both these conditions.  It can be seen that coherence decays
until the first bit-flip occurs, after which it rises, reaches a local
maximum before decohering once again -- with this pattern repeating
between every two bit-flips. It can also be seen that DD recovers most
of the dephasing, that is, $\exp(-\Gamma(t))$ is much closer to unity
than in the uncontrolled evolution, which falls rapidly before
saturating to $\exp(-\Gamma_0(\infty))$. After the first few initial
pulses, the dephasing enters a phase in which the behaviour after the
$(n+1)$-th pulse is approximately the same as the one after the $n$-th
pulse. For $\Delta t=0.1$~ps and $\Delta t=0.2$~ps, the average
dephasing {\it over each cycle} in this `steady-state' phase is very
small, leading to a practical `freezing' of the average decoherence
over a period much longer than the estimated (sub-picosecond) gating
times \cite{irene}.  For $\Delta t=0.3$~ps, however, the increase of
decoherence due to this average dephasing with time is more
noticeable, leading to worse DD performance overall. It can also be
seen that, to minimize the effects of dephasing, any readout on the
qubit should be made half-way between two control pulses.
As it is well known in NMR, this motivates a proper choice of the
observation window, which underlies the Carr-Purcell (CP) sequence
\cite{carr-purcell} and is also discussed in
Ref.~\onlinecite{speranta} in the spin-boson context.

\begin{figure}[t]
\includegraphics*[width=\linewidth]{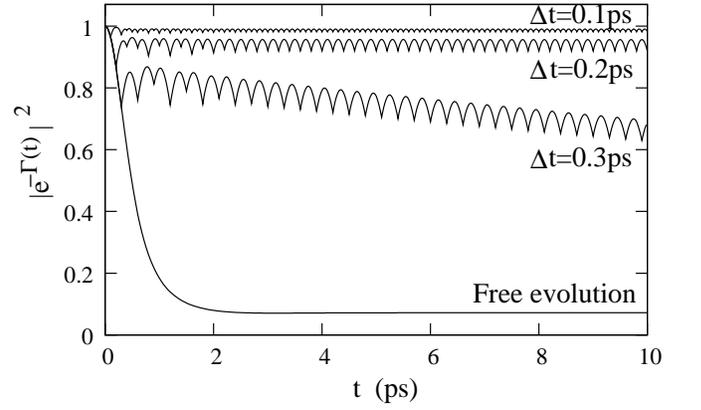}
\caption{$|\exp(-\Gamma(t))|^2$ for the exciton qubit in the presence
of PDD with $\Delta t=0.1$~ps, $\Delta t=0.2$~ps, $\Delta t=0.3$~ps,
compared with the free evolution determined by $\Gamma_0(t)$. }
\label{equidist}
\end{figure}

\subsection{Long-time dynamics: Ohmic versus supraohmic behaviour}
\label{longtime}

A main advantage of the exact representation established in
Eq. (\ref{Gamman-1}) is that it allows detailed quantitative
insight on the controlled dephasing behaviour to be gained.  In
particular, we focus on long-time coherence properties, which have
also received recent attention in view of control-dependent
`saturation' effects observed in the context of spin-bath decoherence
\cite{wenRC} (see also Ref.  \onlinecite{wen09}). We start by
quantifying how the decoherence function in the presence of $n$ pulses
differs between two consecutive control times. Let \be \Delta\Gamma_n
\equiv \Gamma_{n}(t_{n+1})-\Gamma_{n-1}(t_n).
\label{diffd} 
\ee
\noindent 
By using Eq. (\ref{Gamman-1}) we obtain:
\begin{eqnarray}
\Delta\Gamma_n & \hspace*{-1mm}=\hspace*{-1mm}&(-1)^{n}
[\Gamma_0(t_{n+1})-\Gamma_0(t_{n})] +\nonumber\\
&+& 2 \sum_{j=1}^n\Gamma_0(t_{n+1}-t_j)(-1)^{n+j}
\nonumber\\ &\hspace*{-1mm}-\hspace*{-1mm}& 2
\sum_{j=1}^{n-1}\Gamma_0(t_{n}-t_j)(-1)^{j+n}.  
\label{deltagamma}
\end{eqnarray} 

Let now $\Delta\Gamma_n^{\text{PDD}}$ denote the above `differential
dephasing function,' Eq. (\ref{diffd}), specialized to a PDD protocol.
Then, as showed in Appendix \ref{derGaminfty}, the following
asymptotic result holds for an {\em arbitrary dephasing environment}:
\begin{eqnarray}
&&\Delta\Gamma_\infty^{}\equiv \lim_{{ n\to\infty}}
\Delta\Gamma_n^{\text{PDD}}=8\omega_{\text{res}}
\eta\left(\omega_{\text{res}}\right).
\label{deltagammafinal}
\end{eqnarray}
Interestingly, Eq.~(\ref{deltagammafinal}) can be used to describe how
the dephasing function changes between {\em any} two instants
separated by $\Delta t$, for large enough $t$. That is, consider \be
\Delta\Gamma_n^{\text{PDD}} (\tilde{t}) \equiv
\Gamma_{n+1}(\tilde{t}+t_{n+1})-\Gamma_{n}(\tilde{t}+t_n),
\label{gammatilde}
\ee 
\noindent 
where $0\leq \tilde{t}\leq \Delta t$, $t_n=n\Delta t$.
By using Eq.~(\ref{n_eq_nminus1}) we can verify that
$\Delta\Gamma_n^{\text{PDD}} (0)=\Delta\Gamma_n^{\text{PDD}}$.  Then
one may also prove (see Appendix \ref{derGamtildet} for detail) that
\be \Delta\Gamma_n^{\text{PDD}}
(\tilde{t})\stackrel{n>n_{\text{sat}}}{\approx}\Delta\Gamma_\infty,
\label{limitgammatilde}
\ee
\noindent 
where $n_{\text{sat}}\equiv t_{\text{sat}}/\Delta t$ is a sufficiently
large integer defined in the same Appendix.
Eq.~(\ref{limitgammatilde}) shows that the dephasing increment becomes
{\em independent of $n$ and $\tilde{t}$} for $t>t_{\text{sat}}$, that
is, dephasing asymptotically enters a periodic oscillation `in phase'
with the PDD sequence.  Thus, $\Delta\Gamma_\infty$ in
Eq. (\ref{deltagammafinal}) may be used to describe the difference in
dephasing between any two times separated by $\Delta t$ -- in
particular, between consecutive coherence maxima which for
$t>t_{\text{sat}}$ occur at $\tilde{t}\approx\Delta t/2$.  For a
supra-ohmic environment as in the exciton qubit, the convergence of
$\Delta\Gamma_n^{\text{PDD}}$ to $\Delta\Gamma_\infty$,
Eq.~(\ref{deltagammafinal}), is very fast. This is illustrated in
Fig. \ref{dgamma} for two representative values of $\Delta t$.

\begin{figure}[t]
\includegraphics*[width=\linewidth]{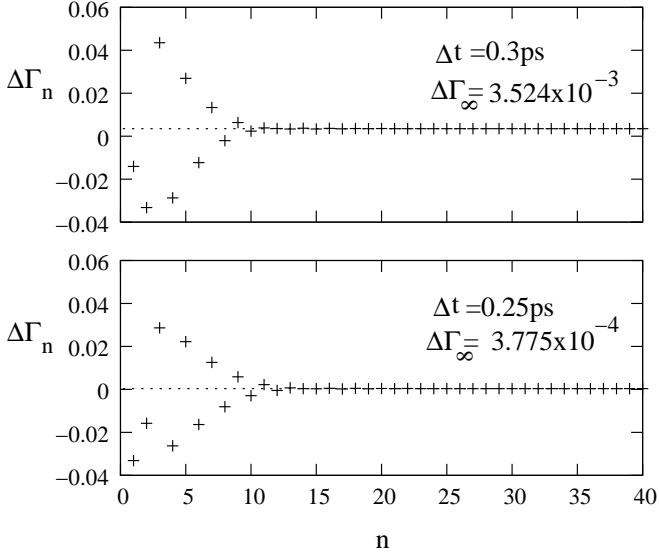}
\caption{Differential dephasing function,
$\Delta\Gamma_n^{\text{PDD}}$, for the exciton qubit under examination
in the presence of PDD with $\Delta t=0.3$~ps (top) and $\Delta
t=0.25$~ps (bottom), calculated from Eq. (\ref{deltagamma}). The
dotted lines show, in each case, the limiting value
$\Delta\Gamma_{\infty}$ given by Eq. (\ref{deltagammafinal}). Notice
that for $n<n_{\text{sat}}$, where $n_{\text{sat}}\sim 15$, the sign
of $\Delta\Gamma_n^{\text{PDD}}$ oscillates, in agreement with
Eqs.~(\ref{TD2}) and (\ref{TI2}). }
\label{dgamma}
\end{figure}

Because $\Delta \Gamma_\infty$ in Eq. (\ref{deltagammafinal}) is
non-zero as long as $\Delta t$ is finite, we can infer that $\Gamma_n$
diverges for fixed $\Delta t$ as $n\rightarrow \infty$.  While this in
principle implies a decay of $\exp(-\Gamma(t))$ to zero under the PDD,
details of the spectral density function (including the nature of the
coupling spectrum and the form of spectral cutoff) become essential
to characterize different dynamical regimes of interest.  In what
follows, we illustrate these features by contrasting ohmic and
supraohmic dephasing environments, and by considering {\em
stroboscopic} sampling, $t_n=2n\Delta t$, in which case explicit
analytic expressions for the PDD `filter function' $|y_{2n}(2n\omega
\Delta t)|^2$ are available.  Specifically, upon combining
Eq.~(11b) of Ref.~\onlinecite{opt} with Eq.~(\ref{n_eq_nminus1})
recovers the well-known result \cite{lorenza,speranta,opt-njp}:
\begin{eqnarray}
\Gamma_{2n}(2n\Delta t)
=\int_0^\infty 4\eta(\omega)\sin^2\left(\omega n \Delta
t\right)\tan^2\Big(\frac{\omega\Delta t}{2}\Big)d\omega.  \nonumber\\
\label{defunc}
\end{eqnarray}

In general, we expect two dominant contributions to the above
integral: the one from small values of $\omega$, where $\eta(\omega)$
is not small, and the one from the region of the resonance,
$\omega\approx\omega_{\text{res}}$, where $|y_{2n}(\omega t)|^2$ may
be large.  First, note that for both a ohmic and supra-ohmic spectral
density, the contributions from the small-$\omega$ region saturate to
a {\it finite} value with time.  For the ohmic case, this is true
irrespective of the fact that the free dephasing dynamics does {\it
not} exhibit a similar long-time saturation.  This behavior is due to
the control term $\tan^2(\omega\Delta t/2)$, which increases the rate
at which the integrand goes to zero as $\omega\to 0$. Second, the
contribution from the $\omega\approx\omega_{\text{res}}$ region is
more or less relevant depending on the form of the spectral cutoff.
Clearly, such `resonating' contributions do not pose a problem in the
limiting situation of an arbitrarily `hard' spectral cutoff of the
form $\Theta(\omega - \omega_c)$ ($\Theta(\:)$ denoting the step
function), since, as remarked earlier, $\omega_{\text{res}}>\omega_c$
in a good DD limit.  For a smooth (`soft') spectral cutoff, the
resonating contribution increases with time and will ultimately be
responsible for the divergence of $\Gamma_{2n}(2n\Delta t)$ as
$n\to\infty$.  In fact, $\Delta\Gamma_\infty$ corresponds precisely to
such a frequency range.  As shown by Eq.~(\ref{limitgammatilde}), we
can approximate $\Delta\Gamma_n\approx\Delta\Gamma_\infty$ for
$t>t_{\text{sat}}$: since at such long times, the contributions to
Eq.~(\ref{defunc}) from small $\omega$ have saturated, dephasing is
indeed dominated from the region around $\omega_{\text{res}}$.  Thus,
for both ohmic and supra-ohmic systems under PDD, the coherence will
eventually decay to zero for large enough times and soft cutoffs.

The above considerations are illustrated in
Fig.~\ref{contributions}, where we plot exact results calculated from
Eq.~(\ref{defunc}) for a representative ohmic spectral density with an
exponential cutoff \cite{lorenza}:
\begin{equation}
I_a(\omega)=F\omega\exp\Big(-\frac{\omega}{\omega_c}\Big),
\label{ohmicsoft}
\end{equation}
In order to highlight the different contributions to the overall
dephasing function, we also explicitly compute and plot the following
quantities: (i) (dotted line)
\begin{eqnarray}
\Gamma_{\text{sm}\omega}(2n\Delta t)=\int_0^{\omega_{\text{res}}/2}
\eta(\omega)|y_{2n}(\omega 2n\Delta t)|^2d\omega,
\label{lowomega}
\end{eqnarray}
which isolates the small-$\omega$ contributions, and (ii) (dashed
line)
\begin{equation}
\Gamma_{\text{res}}(2n\Delta
t)=\int_{\omega_{\text{res}}/2}^{3\omega_{\text{res}}/2}
\eta(\omega)|y_{2n}(\omega 2n\Delta t)|^2d\omega,
\label{rescont}
\end{equation} 
which isolates the contributions from the
$\omega\approx\omega_{\text{res}}$ region.  Three distinct regions may
be identified: an initial drop in coherence due to the low-frequency
modes, until saturation of Eq.~(\ref{lowomega}) occurs at about
$t=\tau_c$; a plateaux region where the contributions from
Eq.~(\ref{rescont}) are not important enough to cause further
decoherence; and a final decay of coherence to zero caused by
increasing contributions from the $\omega\approx\omega_{\text{res}}$
region.  Fig.~\ref{contributions} also compares (bottom panel) the
resonating contributions calculated from Eq.~(\ref{rescont}) with the
asymptotic prediction $\exp(-\Delta\Gamma_\infty t/\Delta t)$ (solid
line), with $\Delta\Gamma_\infty=4.507\times10^{-7}$.  The data
confirm that $\Delta\Gamma_\infty$ does indeed arise from the
resonating contributions as expected, and that as long as the
low-frequency contributions have saturated, $\Delta\Gamma_\infty$ may
be used to accurately describe dephasing under PDD in the long time
limit, that is, $\Delta\Gamma_n\approx\Delta\Gamma_\infty$, for
$t>t_{\text{sat}}$.

\begin{figure}[t]
\includegraphics*[width=7.5cm]{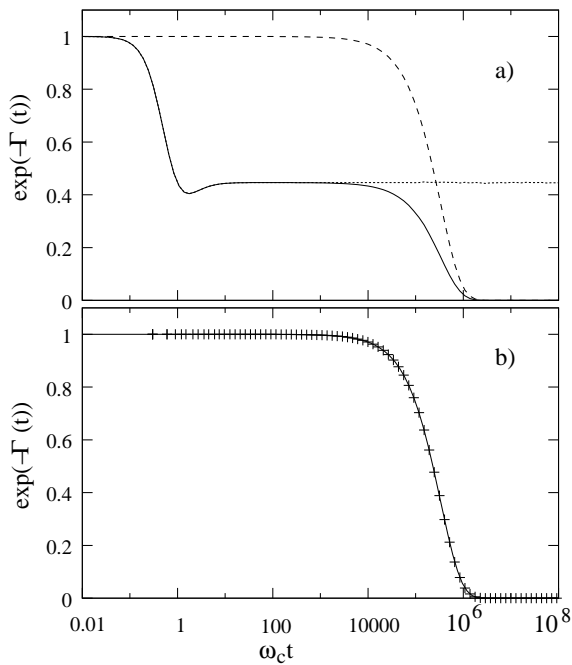}
\caption{Dephasing behavior for an ohmic spectral density with
exponential cutoff as in Eq. (\ref{ohmicsoft}), with $F=0.5$, $\alpha=1/2$, 
$\Delta t=0.0015$, $\omega_c=100$ and $T=100\omega_c$, in units where
$\hbar=k_B=1$.  While stroboscopic sampling is implied, continuous
interpolating lines are used for clarity.  a) Full decoherence
function, $\exp(-\Gamma(t_n))$, Eq.~(\ref{defunc}) (solid line);
low-frequency contribution, $\exp(-\Gamma_{\text{sm}\omega}(t))$,
Eq. (\ref{lowomega}) (dotted line); resonanting contribution,
$\exp(-\Gamma_{\text{res}}(t))$, Eq. (\ref{rescont}) (dashed line)
versus rescaled time $\omega_c t$.
b) Comparison between $\exp(-\Gamma_{\text{res}}(2n\Delta t))$
(Eq.~(\ref{rescont})) (points) and $\exp(-\Delta\Gamma_\infty t/\Delta
t)$ (solid line). }
\label{contributions}
\end{figure}

Additional insight may be gained by examining how the above different
regimes (initial decay, plateaux, final coherence decay) are affected
by the harder or softer spectral cutoff function.  Beside the ohmic
spectral density of Eq. (\ref{ohmicsoft}), consider the following
supra-ohmic spectral densities:
\begin{eqnarray}
I_b(\omega)&=&F\omega^3\exp\Big(-\frac{\omega}{\omega_c}\Big),
\label{supra2}
\\
I_{c}(\omega)&=&F\omega^3\exp\Big(-\frac{\omega^2}{\omega_c^2}\Big),
\label{supra3}
\end{eqnarray}
where, in particular, $I_{c}(\omega)$ has a Gaussian tail, similar to
the excitonic qubit case.   
When comparing $I_b(\omega)$ and $I_{c}(\omega)$ (see
Fig.~\ref{spectComparison}), the harder cut-off due to the Gaussian
tail strongly reduces the value of $\eta(\omega_{\text{res}})$, and
hence greatly increases the duration of the plateaux regime. In fact,
for the set of parameter chosen, our numerics loose the necessary
precision well before the third regime sets on for
$I_{c}(\omega)$. The harder cut-off of the Gaussian case also
decreases $\Gamma_{\text{sm}\omega}$ and, in turn, decreases the
decoherence that occurs before the plateux.

\begin{figure}[t]
\includegraphics*[width=7.5cm]{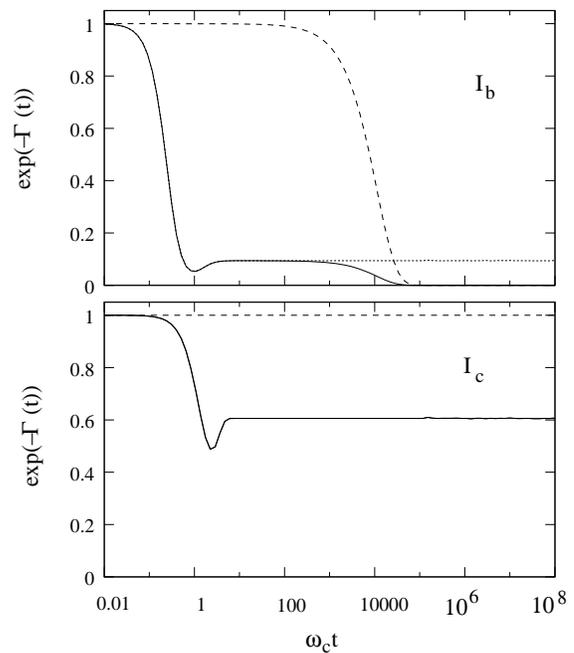}
\caption{Dephasing behavior for supraohmic spectral densities with
different cutoffs, Eqs. (\ref{supra2})-(\ref{supra3}). Notice
that now $F=0.0001$, while all other parameters are as in
Fig.~\ref{contributions}.  $\exp(-\Gamma(t))$ (solid line),
$\exp(-\Gamma_{\text{sm}\omega}(t))$ (dotted line), and
$\exp(-\Gamma_{\text{res}}(t))$ (dashed line) as a function of the
rescaled time $\omega_c t$ for spectral densities $I_b$ (upper
panel), and $I_{c}$ (lower panel), respectively.}
\label{spectComparison}
\end{figure}

\subsection{Short time dynamics}\label{shorttime}

In the previous section we analyzed the dephasing dynamics in the
presence of PDD for $t>t_{\text{sat}}$.  Here, we focus on
$t<t_{sat}$.  The long time regime is entered when
$\Gamma_{n+1}=\Gamma_{n} +\Delta\Gamma_\infty$, and for this to occur
the coherence must oscillate in phase with the DD pulses. However, the
natural response of the coherence after the first PDD pulse is instead
to oscillate with a period of $2\Delta t$ (twice that of PDD pulses,
recall Fig.~\ref{unf}).  This follows from the fact that the first
bit-flip occurs an interval $\Delta t$ after a maximum, $\Gamma_0(0)$,
and for sufficiently small $\Delta t$, the dephasing function is
roughly symmetrical about the control pulse, so the coherence maximum
following the first pulse occurs at $t\approx 2\Delta t$.
The PDD sequence  quickly drives the coherence into phase with it,
 (see Fig.~\ref{unf}), but 
the first few {\it even} bit-flips in PDD occur near the
coherence maxima, and this worsens the performance of the control sequence.
  This may be seen by considering
Eq.~(\ref{Gn-1}) at time $t=t_n+\tilde{t}$, with $0<\tilde{t}\le
\Delta t$. By expanding the first and last terms to first order in
$\tilde{t}$, and considering that $\Gamma_{0} (0)$ is a maximum,
Eq.~(\ref{Gn-1}) rewrites as \be \Gamma_{n}(t)\approx
-\frac{d\Gamma_{n-1}(t_n)}{dt} \tilde{t}+\Gamma_{n-1}(t_n).  \ee
\noindent 
The second term in the above equation is a constant, hence there can
be a coherence peak after the $n$-th control pulses only if the
derivative $ d{\Gamma_{n-1}(t_n)}/dt >0,$  as also pointed out in
Ref.~\onlinecite{speranta}.  In particular, again using
Eq.~(\ref{Gn-1}), we can calculate
$$\frac{d\Gamma_{n}(t_n)}{dt}\approx 
\frac{\Gamma_{n}(t_n+\tilde{t})-\Gamma_{n}(t_n)}{\tilde{t}}
=-\frac{d\Gamma_{n-1}(t_n)}{dt},$$
\noindent 
which shows that the larger the gradient of $\Gamma_{n-1}(t_n)$, the
faster the coherence is retrieved immediately following the $n$th
pulse.  In particular, if $\Gamma_{n-1}(t)$ is locally flat at the
time of the $n$-th pulse, {\it no coherence gain} can occur after
that pulse.

We can see from Fig.~\ref{unf} that as PDD drives the coherence
oscillations into phase with it, $\Delta\Gamma_n$ has alternating sign
for odd and even $n$ (cf. Eqs.~(\ref{TD2}) and (\ref{TI2})).
$\Delta\Gamma_n$ is initially negative for odd $n$ and positive for
even $n$, while its magnitude decreases until a time $t_{\text{av}}$
after which $\Delta\Gamma_n$ becomes positive for odd $n$ and negative
for even $n$, before saturating to
$\Delta\Gamma_n=\Delta\Gamma_\infty$. We see numerically that
$t_{\text{av}}$ is independent of $\Delta t$, with
$t_{\text{av}}\approx0.5$~ps in our case.  Furthermore, we can show
from Eq.~(\ref{deltagammapdd}) that if we consider the times at which
the control pulses occur ($\tilde t=0$), then
\begin{equation}
\Delta\Gamma_n^{\text{PDD}}(0)=\Delta\Gamma^{\text{PDD}}_{n-1}(0)
+(-1)^n\Delta t^2 \Gamma_0''(n\Delta t),
\label{secondderiv}
\end{equation}
where
$$ \frac{d^2 \Gamma_0(n\Delta
t)}{dt^2}\hspace{-0.7mm}=\hspace{-0.7mm}\frac{\Gamma_0((n-1)\Delta
t)-2\Gamma_0(n\Delta t)+\Gamma_0((n+1)\Delta t)}{\Delta t^2}. $$
\noindent  
From this expression we can understand the behaviour of the dephasing
for PDD as the coherence oscillations are driven into phase with the
PDD pulses. As $n$ increases, the sign of the last term in
Eq.~(\ref{secondderiv}) alternates, and its magnitude decreases
as $d \Gamma_0(n\Delta t)/dt$ reaches a maximum before decreasing and
tending to zero (recall the behavior of $\Gamma_0(t)$ in
Fig.~\ref{asymptdec}).  Thus, we can now rigorously define
$t_{\text{av}}$ by the condition $d^2 \Gamma_0(t_{\text{av}})/dt^2=0$,
that is, when the gradient of $\Gamma_0(t)$ is maximum.

\begin{figure}[t]
\includegraphics*[width=\linewidth]{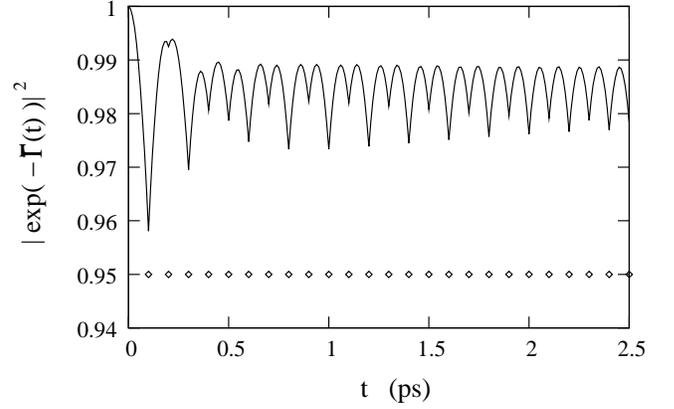}
\caption{Short term dephasing of the exciton qubit under PDD,
with $\Delta t=0.1$~ps.  The diamonds indicate the timing of the PDD
pulses.}
\label{unf}
\end{figure}

\subsection{Practical considerations}

Even if the qubit coherence eventually decays to zero under PDD in our excitonic system, 
for practical purposes we only need to suppress the
dephasing for the qubit lifetime, $T_1$.  From the above discussion,
we can estimate more precisely how short $\Delta t$ must be, in order
for this to happen. For $t=n\Delta t+\Delta t/2>t_{\text{sat}}$, we
can approximate the off-diagonal density matrix element at the maxima
of coherence (where a measurement would be made) as
\begin{equation}
\rho_{01}[(n+1/2)\Delta t]\approx\rho_{01} (0)
e^{-\Gamma_{n_{\text{sat}}}[(n_{\text{sat}} + 1/2 )\Delta t] -
(n-{n_{\text{sat}}})\Delta\Gamma_{\infty}}.
\label{deltagammaevoexact}
\end{equation}

Considering the long-time limit, if $\Delta t$ is sufficiently small
and $n \gg n_{\text{sat}}$, we may neglect the coherence that is lost
whilst $t<t_{\text{sat}}$, and further approximate the dephasing as
\begin{equation}
\rho_{01}[(n+1/2)\Delta t]\approx\rho_{01}(0) e^{-n
\Delta\Gamma_{\infty}}\approx \rho_{01} (0)
e^{-\frac{\Delta\Gamma_{\infty}}{\Delta t} t}.
\label{deltagammaevo}
\end{equation}
Thus, in the long-time limit, we effectively have $1/T_2^{\text{eff}}
= \Delta\Gamma_{\infty}/\Delta t$. A sufficient condition for the
dephasing to be suppressed for the entire qubit lifetime is then
\begin{equation}
T_2^{\text{eff}}= \frac{\Delta t}{\Delta\Gamma_{\infty}} \gtrsim T_1.
\label{t1}
\end{equation}
Fig.~\ref{efficiency} shows $\Delta\Gamma_{\infty}/\Delta t$ as a
function of $\Delta t$ for the exciton qubit under consideration, for
which $1/T_1=1$ ns$^{-1}$.  It can be seen that for $\Delta t \lesssim
0.2$~ps, PDD effectively suppresses dephasing for the entire
lifetime. This is in excellent agreement with our previous results in
Ref. \onlinecite{dephasing}, where we found numerically that $\Delta
t=0.2$~ps leads to efficient PDD, but, in comparison, $\Delta
t=0.3$~ps could only suppress the dephasing for relatively short
times.

\begin{figure}[h]
\includegraphics*[width=7.5cm]{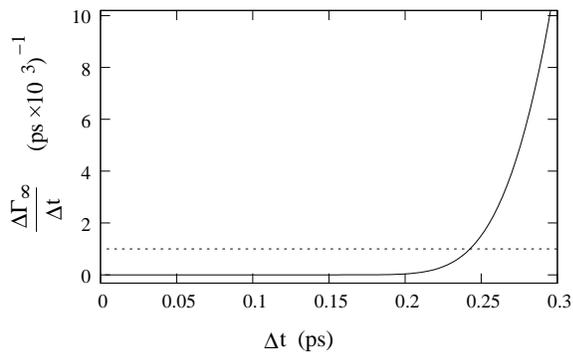}
\caption{Effective long-time coherence decay rate,
$1/T_2^{\text{eff}}=\Delta\Gamma_{\infty}/\Delta t$, Eq. (\ref{t1}),
as a function of $\Delta t$. The dashed line is the qubit inverse
lifetime, $1/T_1=1$ ns$^{-1}$.}
\label{efficiency}
\end{figure}

\section{Comparison of PDD with non-uniform DD schemes}
\label{DDcompare}

Having characterized the performance of the simplest DD scheme, where
the control involves a single time scale $\Delta t$, we proceed to
examine some of the high-level protocols mentioned in the
introduction, which involve {\em non-uniform pulse delays} to a lesser
or greater extent.  While CPDD is both, historically, the most
established approach and, ultimately, one of the most effective, we
defer its discussion until after the analysis of CDD and UDD, since it
turns out that for the supra-ohmic system at hand CPDD naturally
suggests the optimization strategy that will be introduced in
Sec. \ref{Constrained}.

\subsection{Concatenated decoupling}

Instead of repeating the basic control cycle given in
Eq.~(\ref{cycle}), CDD recursively concatenates it within itself.  Let
$S_\ell$ denote the sequence corresponding to the $\ell$-th level of
concatenation, as given in Table \ref{concatenations}.

For a qubit undergoing arbitrary decoherence, CDD with a `universal
decoupling' cycle given, for instance, by $\Delta t X\Delta t Z\Delta
t X \Delta t Z,$ has been shown \cite{khodjasteh} to significantly
outperform PDD in the limit $\Delta t \to 0$. However, for purely
dephasing systems for which $\Delta t$ has a {\em finite} lower limit,
and for {\em single-axis} protocols constructed out of the basic cycle
in Eq. (\ref{cycle}), the advantages of CDD are largely lost, and PDD
may be more efficient \cite{remark2}.  While different ways for
comparing different DD protocols can be considered
\cite{khodjasteh,nave,wen}, we shall focus here on comparing the
efficiency of PDD and CDD at ensuring dephasing-protected storage of
the exciton qubit for a {\em fixed} time $T_{\text{storage}}$. In
particular, for our calculations we choose $T_{\text{storage}}=10$~ps.
This time is appropriate given the typical gating time for
exciton-based QIP, which is of the order of 1 ps\cite{irene}.

\begin{table}[t]
\begin{center} {\footnotesize
\begin{tabular}{ccc}
\hline Sequence & Pulse Timing\\ \hline $S_0$& \text{Free}$(\Delta
t)$\\ $S_1$& $X \Delta t X \Delta t$\\ $S_2$& $X [X \Delta t X \Delta
t] X [X \Delta t X \Delta t] = \Delta t X \Delta t \Delta t X \Delta
t$\\ $S_3$& $X[\Delta t X \Delta t \Delta t X \Delta t]X[\Delta t X
\Delta t \Delta t X \Delta t] $\\ $\vdots$ & $\vdots$\\ $S_\ell$& $X
S_{\ell-1} X S_{\ell-1}$\\ \hline
\end{tabular} }
\end{center}
\caption{Concatenated pulses sequences for a purely dephasing
single-qubit interaction. Time ordering is from right to left. }
\label{concatenations}
\end{table} 

\subsubsection{Single CDD cycle}
\label{CDDone}

Given $T_{\text{storage}}$ and the presence of a physical constraint
on $\Delta t$, a first way to exploit CDD is to identify a minimum
concatenation level, $\ell^\ast$, for which the length of the
corresponding sequence, $T_{\ell^\ast}=2^{\ell^\ast}\Delta t$, exceeds
$T_{\text{storage}}$. For a given $\Delta t$, increasing $\ell$ beyond
this point would not modify the results because the pulse timings over
$T_{\text{storage}}$ would be unchanged.   (see Table
\ref{concatenations}).
Figure~\ref{cddvspdd} compares CDD and PDD for storage of an exciton
qubit for different $\Delta t$.  
As expected from the general analysis of Ref.~\onlinecite{khodjasteh},
the efficiency of CDD increases with decreasing $\Delta t$.  However,
in the range of values under exploration, and with readout effected at
the maxima of the coherence curve, CDD is found to be more efficient
than PDD only if $\Delta t\lesssim 0.036$~ps.  The latter time scale
is substantially smaller than physically allowed in our system.

\begin{figure}[t]
\includegraphics*[width=\linewidth]{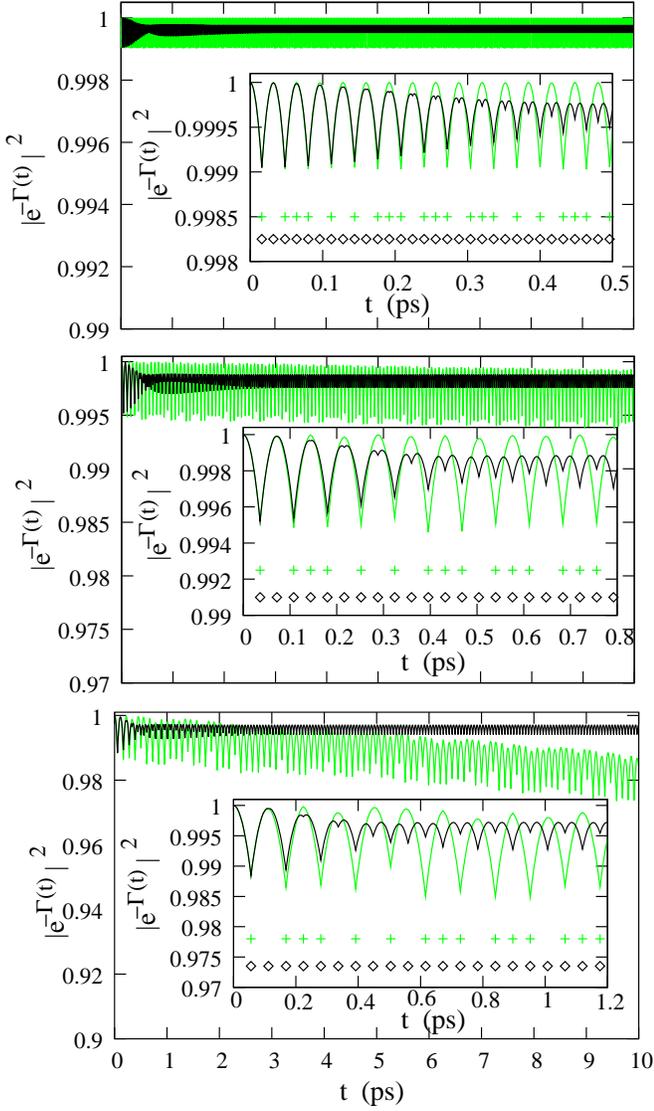}
\caption{(Color online) $|\exp(-\Gamma(t))|^2$ for PDD (black line)
compared with CDD (light line) for $\Delta t=0.016$~ps ($\ell^\ast
=10$, top), $\Delta t=0.036$~ps ($\ell^\ast =9$, middle), and $\Delta
t=0.055$~ps ($\ell^\ast =8$, bottom). Insets: Close-ups of the same
evolutions at short times; the pulse timings are indicated as well,
with crosses (CDD) and diamonds (PDD).  }
\label{cddvspdd}
\end{figure}

We can understand the possible advantage of CDD by comparing it with
the long- and short-time behaviour of PDD (Secs. \ref{longtime}
and \ref{shorttime} respectively).  Eq.~(\ref{deltagammaevoexact})
shows that the long-time performance of the protocol depends on
$\Delta\Gamma_{\infty}$, and $\Gamma_{n_{\text{sat}}}[
(n_{\text{sat}}+1/2)\Delta t]$. For very small $\Delta t$  (hence
small $\Delta\Gamma_{\infty}$), PDD is not the most efficient scheme
because it leads to a value of $\Gamma_{n_{\text{sat}}}
[(n_{\text{sat}}+1/2)\Delta t]$ which may be greater than for other
pulse sequences, due to the initially `out of phase' pulses.  In the
regime where CDD outperforms PDD (very small $\Delta t$), the
contributions to dephasing from around $\omega=\omega_{\text{res}}$
(see Sec.~\ref{longtime}) are negligible for both sequences over
$T_{\text{storage}}$, since for $t>t_{\text{sat}}$ both
sequences preserve the maxima of coherence very close to the value
$\exp(-\Gamma(t_{\text{sat}}^{\text{max}}))$ corresponding to the time
$t_{\text{sat}}^{\text{max}}$ of the first maximum that follows
$t_{\text{sat}}$.
The advantage of CDD (if any) comes from the different behaviour of
the dephasing over the first few control pulses, that is, up to
$t=t_{\text{sat}}$. The timing of the pulses in the CDD sequence are
similar to those of PDD, but with {\it fewer} pulses at the instants
where the even pulses occur in PDD.  These `missing' pulses are those
which would occur near the maxima of coherence in the initial stages
of the sequence (see insets of Fig.~\ref{cddvspdd}), that is, the ones
responsible for decreasing the coherence maxima while
$t<t_{\text{sat}}$ in PDD (Sec.~\ref{shorttime}). These `missing'
pulses allow the dephasing to maintain its natural response frequency
after the first bit-flip, and no loss of dephasing is needed to change
the rate of the oscillations of coherence. Therefore,
$\Gamma^{\text{CDD}}(t_{\text{sat}}^{\text{max}})
>\Gamma^{\text{PDD}}(t_{\text{sat}}^{\text{max}})$, and for
$t<T_{\text{storage}}$,
$\Gamma(t)\approx\Gamma(t_{\text{sat}}^{\text{max}})$ for both PDD and
CDD in the limit of sufficiently small $\Delta t$.

While the above explains why CDD may outperform PDD, as soon
as $\Delta t$ is long enough such that $\Delta\Gamma_{\infty}$ is
significant over $T_{\text{storage}}$, PDD becomes the most efficient
sequence. The period of the coherence oscillations for CDD is twice
the one for the PDD sequence corresponding to the same $\Delta t$ (see
insets in Fig.~\ref{cddvspdd}), resulting in faster dephasing at long
times $t$ for CDD.

\subsubsection{Periodic repetition of CDD cycles}
\label{PCDDsec}

A different use of CDD consists in truncating concatenation at a
fixed level and periodically repeating the resulting `supercycle',
constructed from Table.~\ref{concatenations}.  For instance,
truncation at $\ell =2$ results in our purely dephasing case in a
cycle of length $4 \Delta t$, which is identical in structure to a CP
cycle (see Sec. \ref{carr}), and whose periodic repetition we term
PCDD$_2$.  For a single qubit undergoing arbitrary decoherence, the
corresponding PCDD$_2$ protocol (constructed from a $16$-pulse base
cycle) has been shown to be the best performer in suppressing the
effects of a quantum spin bath \cite{wen,wenRC,wenJMO}.

Figure~\ref{PCDD} shows a comparison of PDD and PCDD$_\ell$ protocols
for $\ell =2,3$, for the shortest pulse separation compatible with the
exciton qubit constraint, $\Delta t=0.1$~ps.  One can infer that, for
the $\Delta t$ and $T_{\text{storage}}$ values considered, PCDD$_\ell$
performs better (that is, displays higher coherence maxima) than PDD
for $\ell =2$, but worse for $\ell =3$.
The difference between PCDD$_2$ and PCDD$_3$ may be understood as
a consequence of the fact that in terms of a Magnus expansion
\cite{magnus}, concatenated cycles with even $\ell$ are
time-symmetric, thus cancel the interaction with the phonon bath up to
(at least) the second order.  Over the time period shown, PCDD$_2$
also outperforms standard PDD (see Fig.~\ref{PCDD}, upper panel).
However, the coherence oscillations for PCDD$_2$ occur over a period
of $2\Delta t$ since, after the initial pulse, the sequence is 
equivalent to PDD with a base time interval of $2\Delta t$.
Therefore, we expect PDD to be more efficient for long storage times,
as PCDD$_2$ will yield a larger $\Delta\Gamma_{\infty}$ than a PDD
sequence characterized by $\Delta t$, hence worse asymptotic
performance.

\begin{figure}
\includegraphics*[width=\linewidth]{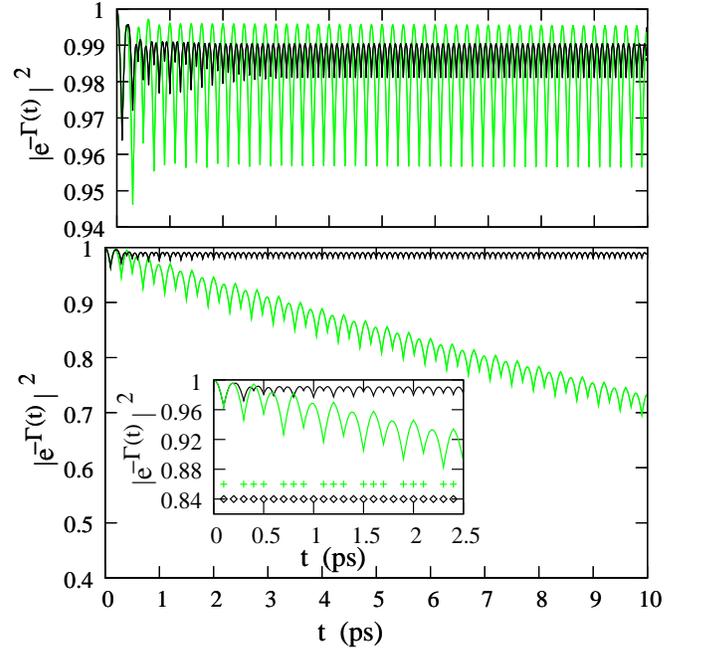}
\caption{(Color online) Comparison of PDD (dark line) and PCDD (light
line) protocols with $\Delta t=0.1$~ps.  Top: Second-level
concatenated cycle, PCDD$_2$. Bottom: Third-level concatenated cycle,
PCDD$_3$. Inset: Zoom over the initial part of the time window with
timings of the pulse sequences explicitly indicated (diamonds for PDD
and crosses for PCDD$_3$).}
\label{PCDD}
\end{figure}

\subsection{Uhrig decoupling}

We now assess the limitations of the optimal sequence proposed
by Uhrig \cite{opt} when significant restrictions on $\Delta t$ are in
place.  In UDD, consecutive pulses are spaced according to
\begin{equation}
\delta_j=\sin^2 \Big(\frac{\pi j}{2n+2}\Big) ,
\label{optimised}
\end{equation}
which implies, in particular, closely spaced pulses at the beginning
and the end of the evolution period. Such a control sequence strongly
suppresses the dephasing for a storage time of the order of \cite{opt}
\begin{equation}
t_{\text{UDD}}\approx(n+1)\frac{\tau_c}{2\pi},
\label{optimisedtime}
\end{equation}
where $\tau_c$ denotes, as before (Sec. \ref{pdd}), the relevant 
bath correlation time.  As mentioned, with $\hbar\omega_c\approx
2$~meV, this corresponds to $\tau_c \approx 2.06$ ps.
Beyond $t_{\text{UDD}}$, the efficiency of UDD falls rapidly. From
Eq.~(\ref{optimisedtime}), we find that for UDD to efficiently protect
the exciton qubit over $T_{\text{storage}} \approx 10$~ps, $n$ must be
on the order of $100$. Fig. \ref{urighdomain} shows the resulting UDD
performance as $n$ is decreased.  It can be seen that as $ n \lesssim
100$, the advantage of UDD is rapidly lost.

\begin{figure}[t]
\includegraphics*[width=\linewidth]{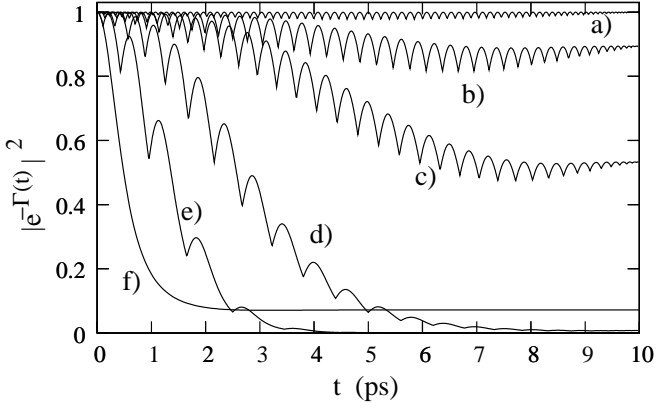}
\caption{$|\exp(-\Gamma(t))|^2$ for the exciton qubit in the presence
of UDD with a) $100$ pulses, b) $50$ pulses, c) $40$ pulses, d) $25$
pulses, e) $14$ pulses corresponding to the best allowed sequence for
the case of the exciton qubit.  For comparison, f) shows the free
evolution.  }
\label{urighdomain}
\end{figure}

For our QD system, however, the main physical limitation is on the
time delay between pulses.  The shortest interval between control
pulses in UDD, $\Delta t_{\text{min}}^{\text{UDD}}$, is before the
first pulse, and after the last pulse. From Eq.~(\ref{optimised}) we
see that such a sequence with $n =100$ pulses over a period of
$T_{\text{storage}}=10$~ps corresponds to $\Delta
t_{\text{min}}^{\text{UDD}}=2.4$x$10^{-3}$ ps, which is roughly {\it
two orders of magnitude} less than that allowed by the physical
constraints for the exciton qubit in question.  Even for a sequence
consisting of $n=40$ pulses only (for which the efficiency is already
poor as shown in Fig. \ref{urighdomain}, curve c)), $\Delta
t_{\text{min}}^{\text{UDD}}=1.5$x$10^{-2}$ ps, which is still an order
of magnitude shorter than allowed.

To respect the physical constraints, one may estimaate that
allowed UDD sequences should have a number of pulses $n \lesssim 14$
within the intended $T_{\text{storage}}=10$~ps.  Such a sequence
corresponds to curve e) in Fig. \ref{urighdomain}. It is then clear
that any UDD sequence compatible with our physical constraints is
outperformed by the best allowed PDD sequence which would preserve a
coherence close to 1 for the same time window (see
Fig.~\ref{equidist}).  Fig.~\ref{urighdomain} also shows that any
constrained UDD sequence performing like curve d) or worse would
increase the dephasing compared with the free evolution, that is, 
would result in decoherence acceleration.  The reason for the
shortfalls of UDD in our setting stems from the large spread of
the control intervals $(t_i-t_{i-1})$.  If we impose a lower bound on
the minimum time interval, other intervals must take up a
considerable proportion of the total evolution time.  This
places a relatively large restriction on how many pulses may be used
within a given storage period, and eventually results in large
amounts of dephasing during the long time delays in which no pulses
occur.

\subsection{Carr-Purcell decoupling}
\label{carr}

We now focus on analyzing more closely CPDD, which results from the
periodic repetition of a CP cycle of the form \cite{carr-purcell}
\begin{equation}
\Delta t^{\text{CP}}X \,2\Delta t^{\text{CP}} X \Delta t^{\text{CP}}.
\label{CPcycle}
\end{equation}
This also corresponds, as noted, to PCDD$_2$ with $\Delta
t^{\text{CP}}=\Delta t$ (cf.  Table \ref{concatenations}).
Specifically, we are interested in comparing a PDD sequence with a
CPDD having the {\em same cycle time}, $T_c=2\Delta t$, thus $\Delta
t^{\text{CP}}=\Delta t/2$: though the corresponding pulse time
interval may not be allowed by the physical constraints we are
considering, this study will pave the way to be the analysis to be
developed in the next section.

Basically, CPDD may be viewed as a PDD protocol where pulses are
uniformly spaced by $2\Delta t^{\text{CP}}$, except that the sequence
is displaced forward by $t_1=\Delta t/2$, the time at which the first
pulse is applied.  As a consequence of the symmetry of the control
propagator in Eq. (\ref{CPcycle}) with respect to the cycle mid-point,
it is well known \cite{NMR} that CPDD is a second-order protocol as
compared to standard (asymmetric) PDD, with leading corrections of
order $T_c^3$.  Using the exact representation established in
Eq.~(\ref{Gamman-1}), we will now assess the extent to which CPDD
improves over PDD for a purely dephasing system, and gain insight into
asymptotic properties.

We begin by determining the dephasing half-way between consecutive
control pulses for the case of PDD. Using Eq.~(\ref{Gamman-1}), we
find
\begin{eqnarray}
&&\Gamma^{\text{PDD}}_n\Big[t=\Big(n+\frac{1}{2}\Big)\Delta
t\Big]=2\sum_{m=1}^n(-1)^{m+1}\Gamma_0(m\Delta t)\nonumber\\
&&+4\sum_{m=2}^n\sum_{j<m}\Gamma_0((m-j)\Delta
t)(-1)^{m-1+j}\nonumber\\
&&+2\sum_{m=1}^n(-1)^{m+n}\Gamma_0\Big[\Big(n+\frac{1}{2}-m\Big)\Delta
t\Big]\nonumber\\
&&+(-1)^n\Gamma_0\Big[\Big(n+\frac{1}{2}\Big)\Delta t\Big],
\end{eqnarray}
which we may rewrite as
\begin{eqnarray}
&&\Gamma_n^{\text{PDD}}\Big[\Big(n+\frac{1}{2}\Big)\Delta
t\Big]=2\sum_{m=1}^n(-1)^{m+1}\Gamma_0(m\Delta t)\nonumber\\
&&+4\sum_{m=2}^n\sum_{j<m}\Gamma_0((m-j)\Delta
t)(-1)^{m-1+j}\nonumber\\
&&+2\sum_{k=1}^n(-1)^{k+1}\Gamma_0\Big[\Big(k-\frac{1}{2}\Big)\Delta
t\Big]\nonumber\\ &&+(-1)^n\Gamma_0\Big[\Big(n+\frac{1}{2}\Big)\Delta
t\Big],
\label{pddhalfway}
\end{eqnarray}
where $k=n-m+1$.  Similarly, by using Eq.~(\ref{Gamman-1}), we may
also determine the dephasing for CPDD with $\Delta
t^{\text{CP}}=\Delta t/2$ and $t_i=(i-1/2)\Delta t$, that is,
\begin{equation}
\Gamma_n^{\text{PDD}}[ (n+1/2)\Delta t ]=\Gamma_n^{\text{CPDD}}[ (n+
1/2 ) \Delta t].
\label{compare}
\end{equation}
This {\em exact} result is illustrated in Fig.~\ref{maxmin}, where we
plot the dephasing behaviour under PDD and CPDD for the exciton qubit
with $\Delta t=0.1$~ps. As predicted by Eq.~(\ref{compare}), the
coherence in the presence of each sequence is equal at times
$t=(n+1/2)\Delta t$. Interestingly, for $t>t_{\text{sat}}$,
$\Gamma^{\text{PDD}}_n [(n+1/2) \Delta]$ are local maxima of coherence
whereas $\Gamma_n^{\text{CPDD}} [(n+ 1/2)\Delta t]$ are local minima,
proving CPDD to be much more efficient than PDD provided that the time
of the first pulse is allowed to be $t_1=0.05$ ps.

\begin{figure}[hb]
\includegraphics*[width=\linewidth]{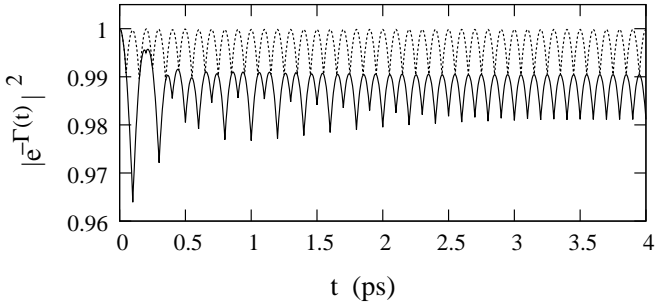}
\caption{Comparison of CPDD (dotted line) with $\Delta t^{CP}=0.05$~ps
and PDD (solid line) with $\Delta t=0.1$~ps for the exciton system.  }
\label{maxmin}
\end{figure}

\section{Towards optimized sequences in the presence of pulse 
timing constraints}
\label{Constrained}

Building on the understanding gained from the comparison between
different protocols in Sec. \ref{DDcompare}, we now specifically aim
to optimize DD performance for a bosonic dephasing environment when
pulses are
 subject to a minimum pulse-delay constraint.

The basic observation is to note that if after an initial arbitrary
pulse sequence, PDD is turned on at a time $t_{\text{PDD}}$, then for
$t>t_{\text{PDD}}+t_{\text{sat}}$, we have $\Gamma_{n+1}(t+\Delta t)-
\Gamma_{n}(t)= \Delta\Gamma_{\infty}$ (recall
Eq.~(\ref{limitgammatilde})).  This naturally suggests an {\em
interpolated DD} approach, where an initial sequence is chosen to
minimize $\Gamma(t_{\text{sat}}^{\text{max}})$, whilst transforming
the oscillations of coherence into phase with a PDD sequence to be
turned on immediately afterwards.  Interestingly, a similar philosophy
has been invoked to optimally merge deterministic and randomized DD
methods to enhance performance over the entire storage time
\cite{LeaJMO}.  In our case, CPDD is indeed the simplest example of
this interpolation: as already noted, CPDD can be thought of as a PDD
sequence applied at $t_{\text{PDD}}=\Delta t/2+\Delta t$, following a
preparatory sequence consisting of a single pulse at $t=\Delta t/2$.

Unfortunately, standard CPDD is not allowed in our sysyem due to the
physical constraint: the time interval between pulses in the initial
sequence is smaller than the minimum allowed $\Delta t$ which
characterizes the subsequent PDD sequence.  Simply using a CPDD
sequence which does not break the time constraint is clearly not
optimal.  If the smallest allowed pulse interval is $\Delta
t_{\text{min}}$, then the best CPDD sequence consists of periodic
repetitions of a CPDD cycle with $\Delta t^{\text{CP}}=\Delta
t_{\text{min}}$, and the most efficient allowed PDD sequence is
repetitions of $X\Delta t_{\text{min}}X\Delta t_{\text{min}}$. Since
CPDD cancels the terms in the Magnus expansion up to to the second
order, over the first few repetitions it performs much better than
PDD, which only cancels them up to the first order. However, for
longer times the effects due to the higher-order Magnus corrections
accumulate, and they turn out to do so more favorably for PDD. This
manifests itself in a smaller $\Delta\Gamma_{\infty}$ for PDD than for
the best allowed CPDD protocol.  As shown by Eqs.~(\ref{TI}) and
(\ref{limitTD}), the coherence oscillations are independent of the
timing of any pulses applied before $t-t_{\text{sat}}$. Therefore,
CPDD can be treated as a PDD sequence with $\Delta t=2\Delta
t^{\text{\text{CP}}}$ for $t>t_{\text{sat}}$. This justifies defining
a $\Delta\Gamma_\infty$ for a CPDD sequence.

Physically, what is needed is a different initial sequence that
efficiently `engineers' the transition of the coherence oscillations
-- from the natural response frequency determined by the first
bit-flip to the frequency of the following PPD sequence. To accomplish
this, we propose to use {\em CP cycles with varying $\Delta
t^{\text{CP}}$}.  That is, we define such an interpolated sequence by
letting the $i$th cycle to be characterized by a pulse delay $\Delta
t^{\text{CP}}_i$, and begin immediately after the previous cycle at
$t_i=t_{i-1}+4\Delta t^{\text{CP}}_{i-1}$.  The analysis of the
resulting averaging properties may be carried out by adapting the
derivation of Ref.~\onlinecite{khodjasteh} to the pure dephasing
bosonic setting of Eq. (\ref{arbhamiltonian}).  While the detail
of the calculations are included in Appendix C, the result is that,
similar to standard CP, the proposed DD sequence still cancels the
terms in the Magnus expansion up to the second order.  Therefore, the
interpolated scheme does not only perform well for small $t$, but also
quickly results in pulses uniformly separated by $\Delta
t_{\text{min}}$ -- resulting in a small $\Delta\Gamma_{\infty}$, hence
high performance for long storage times.

The simplest way to generate a good interpolated DD sequence is to
apply a CP cycle with $\Delta t^{\text{CP}}=\Delta t_{\text{min}}$,
followed by periodic repetitions of one with $\Delta
t^{\text{CP}}=\Delta t_{\text{min}}/2$.  The sequence is then given by
\begin{eqnarray}
t_1&=&\Delta t_{\text{min}},\nonumber\\ t_2&=&3\Delta
t_{\text{min}},\nonumber\\ t_3&=&3\Delta
t_{\text{min}}+\frac{3}{2}\Delta t_{\text{min}},\nonumber\\
t_{i}&=&t_{i-1}+\Delta t_{\text{min}}, \;\;\; i >3.
\label{short}
\end{eqnarray}
We compare this sequence with standard PDD with $\Delta t=\Delta
t_{\text{min}}$ in Fig.~\ref{graph1}, upper panel.  One clearly sees
that the sequence in Eq.~(\ref{short}) is more efficient.

By construction, the first two CP cycles in the above sequence play
the role of modifying the frequency of the dephasing oscillations in
such a way that they are brought in phase with the following repeated
cycles. We can perform this process more smoothly by gradually
reducing $\Delta t^{\text{CP}}$ from $\Delta t_{\text{min}}$ to
$\Delta t_{\text{min}}/2$ over more than a control cycle. Though for
very small $\Delta t_{\text{min}}$ the two cases would be equivalent,
for systems such as the exciton qubit where the time restrictions are
relatively severe, the smoother transition sequence may decouple the
qubit more efficiently. Such a modified sequence may be implemented by
applying CP cycles with decreasing $\Delta t^{CP}$, that is,
\begin{eqnarray}
\Delta t^{\text{CP}}_i = \left\{ \begin{array}{ll}
   \Delta t_{\text{min}}, & \;i=1 \\
   \Delta t_{\text{min}}-(i-1)\Delta_2, 
    & \; 1< i \leq i_{\text{PDD}} , \\ 
{\Delta t_{\text{min}}}/{2}, & \; i >  i_{\text{PDD}},
\end{array}\right.
\label{long}
\end{eqnarray}
where $ i_{\text{PDD}}={\Delta t_{\text{min}} }/(2 \Delta_2)$ and
$\Delta_2$ is an arbitrary time defined such that $i_{\text{PDD}}$ is
an integer. The greater $i_{\text{PDD}}$, the longer the time over
which the the decreasing length cycles are applied.  Alternatively, we
may describe the above sequence in terms of the pulse times:
\begin{eqnarray}
t_i & = & \Delta t_{\text{min}}+(i-1)
2\Delta t_{\text{min}} \nonumber \\ 
&& -\frac{(i-1)(i-2)}{2}\Delta_2, \;\;\;\;
i <\frac{\Delta t_{\text{min}}}{\Delta_2} -2, \nonumber \\ 
t_i &= & t_{i-1}+\Delta t_{\text{min}},
\;\;\;\; i \ge \frac{\Delta t_{\text{min}}}{\Delta_2}- 2.
\end{eqnarray}
The above modified sequence is compared to PDD with $\Delta t=\Delta
t_{\text{min}}$ in Fig.~\ref{graph1}, lower panel.  As before, we see
that DD with varying CP cycles outperform PDD. Furthermore, it can be
seen that there is an improvement over the more abrupt sequence
described by Eq.~(\ref{short}).

\begin{figure}[t]
\includegraphics[width=\linewidth]{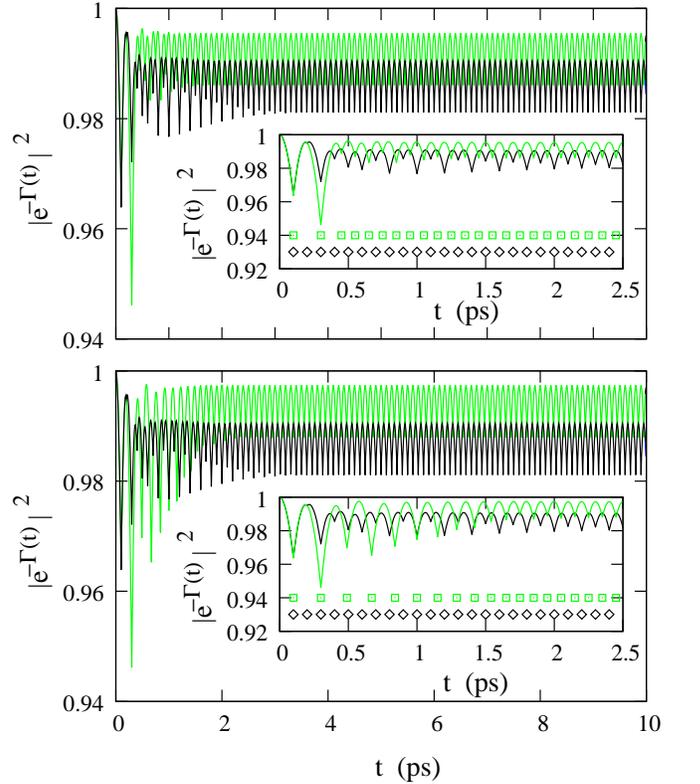}
\caption{(Color online) Comparison of PDD (dark line) with
interpolated DD sequences of varying CP cycles (light line), for
$\Delta t_{\text{min}}=0.1$~ps. Top: Sequence given by
Eq.~(\ref{short}). Bottom: Sequence given by Eq.~(\ref{long}), with
$\Delta_2=0.01$~ps. Inset: Zoom over the initial part of the time
window with timings of the pulse sequences explicitly indicated
(diamonds for PDD and squares for the modified sequences).}
\label{graph1}
\end{figure}

In Sec.~\ref{PCDDsec}, we showed that a constrained CPDD sequence
could outperform PDD over time scales of the order of $10$~ps (see top
panel of Fig.~\ref{PCDD} for PCDD$_2$) even though, for longer times,
the smaller $\Delta\Gamma_{\infty}$ for PDD would eventually make it
more efficient than CPDD.  In Fig.~\ref{carcomp}, we further compare
CPDD with the interpolated sequence given in Eq.~(\ref{long}).  The
latter is found to be slightly more efficient than the best allowed
CPDD sequence over short time scales.  Furthermore, because of the
smaller $\Delta \Gamma_{\infty}$, as time progresses it will also
outperform CPDD asymptotically.  A main advantage of the sequence
given in Eq.~(\ref{long}), however, is that it not only leads to
higher maxima than CPDD, but also, after the first few pulses, to a
{\it much smaller coherence oscillation amplitude}. This reflects the
fact that the oscillation period has been tuned to the minimum allowed
time interval $\Delta t_{\text{min}}$. In this respect, the
performance of the sequence in Eq.~(\ref{long}) is more robust against
the precise readout times, or,
equivalently, a readout offset error relative to the coherence maxima
would not significantly affect the coherence recovered using this
sequence.

\begin{figure}[t]
\includegraphics[width=\linewidth]{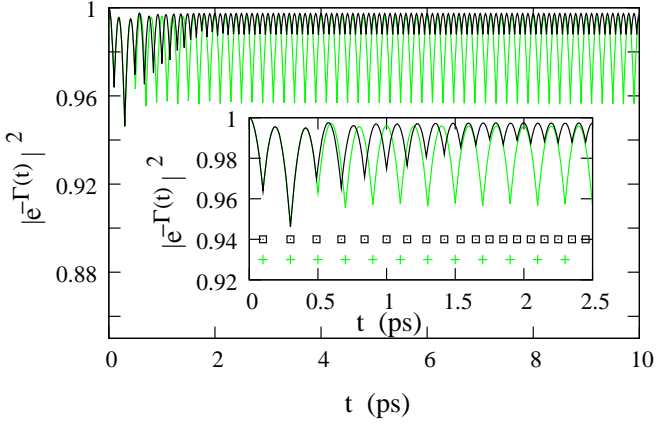}
\caption{(Color online) Comparison of the best allowed CPDD sequence
(light line) with a sequence of CP cycles of decreasing length (dark
line) as described in Eq.~(\ref{long}) with $\Delta_2=0.01$~ps and for
$\Delta t_{\text{min}}=0.1$~ps. Inset: Zoom over the initial part of
the time window with timings of the pulse sequences explicitly
indicated (squares for the modified sequence and crosses for CPDD,
respectively).}
\label{carcomp}
\end{figure}

\section{Conclusion and outlook}

We have investigated the ability of DD to inhibit decoherence of a single qubit coupled to a purely dephasing bosonic environment, by comparing the performance of low-level periodic DD schemes based on uniform pulse separations to higher-level non-uniform DD schemes.

For arbitrary spectral density functions characterizing different
dephasing environments, we have derived an exact representation of the controlled dynamics
available for instantaneous pulses, and exact relationships for the
asymptotic dynamics in the case of PDD.  Building on these results, we
have shown that a main weakness of PDD is due to the oscillation of
coherence following the first bit-flip being out of phase with the
rest of the sequence.  This has naturally suggested the application of
a suitably engineered preparatory sequence as a strategy to enhance DD
efficiency, by bringing the coherence oscillations into phase with a
subsequent PDD sequence.  The resulting `interpolated' DD protocols
are found to be especially efficient for physical systems where the
mimimum time interval between control pulses is strongly constrained.
For such systems, DD protocols like concatenated or Uhrig DD, which
are designed to achieve peak performance when the asymptotic regime of
arbitrarily small pulse separations is fully accessible, tend to
largely lose their advantages.  

For the excitonic dephasing environment of interest,
in particular, we have shown how a sequence of Carr-Purcell cycles
with suitably chosen (analytically generated) time delays provides a
very efficient DD protocol for realistic QD parameters and qubit
storage times.  Our process of constructing a DD sequence under
which the coherence oscillates asymptotically with the minimum period
allowed by the physical contraints offers, as by-product, the
advantage of a {\em significantly smaller coherence oscillation
amplitude}, relative to constrained PCDD or UDD sequences. This makes
the proposed interpolated sequence more {\it robust} against readout.

While our analytically-designed interpolated DD protocol might be 
compelling in its simplicity, identifying DD schemes that are
guaranteed to yield optimal performance subject to non-trivial timing
constraints appears as an interesting control-theoretic problem for
further investigation.  Revisiting the local numerical optimization
approach recently proposed in Ref. \onlinecite{nist2} in a {\em
constrained minimization} perspective might offer a concrete starting
point in this respect.  Likewise, the investigation of dynamical
error-control schemes based on bounded-strength `Eulerian' DD
\cite{Viola:2003:037901}, along with the recently proposed extension
to decoherence-protected quantum gates \cite{DCGs,DCGLong}, might
prove especially fruitful for exciton qubits, in view of the reduced
control overheads associated with purely dephasing environments.
Lastly, an interesting general question is to what extent exact
representations of the controlled coherence dynamics in terms of the
uncontrolled one may exist for an arbitrary purely dephasing error
model, allowing, for instance, exact insight into the long-time
controlled dynamics to be gained.


\acknowledgments

It is a pleasure to thank Kaveh Khodjasteh for a critical reading of
the manuscript.  L.V. gratefully acknowledges partial support from the
National Science Foundation through Grants No. PHY-0555417 and
No. PHY-0903727, and from the Department of Energy, Basic Energy
Sciences, under Contract No. DE-AC02-07CH11358.


\appendix
\section{derivation of $\Delta\Gamma_\infty$\label{derGaminfty}}

For PDD, $t_n=n\Delta t$, using this expression and recasting the sums
in Eq.~(\ref{deltagamma}) in terms of $k=n-j$, we can write
\begin{eqnarray}
\Delta\Gamma_n^{\text{PDD}} &=&(-1)^{n}\Gamma_0((n+1)\Delta t)
- 3\Gamma_0(n\Delta t)(-1)^{n} \nonumber\\
&-&4\sum_{k=1}^{n-1}\Gamma_0(k\Delta t)(-1)^{k}.
\label{deltagammapdd}
\end{eqnarray}
Using Eq.~(\ref{gammaneta}) for $n=0$ and extending the sum to include $k=0$,
Eq.~(\ref{deltagammapdd}) becomes
\begin{eqnarray}
\Delta\Gamma_n^{\text{PDD}} &=&-4\int_0^\infty\eta(\omega)d\omega
+(-1)^n\int_0^\infty6\eta(\omega)\cos(\omega n\Delta
t)d\omega\nonumber\\ 
&& -(-1)^n\int_0^\infty2\eta(\omega)\cos(\omega
(n+1)\Delta t)d\omega \nonumber \\
&& +8\int^\infty_0\eta(\omega)\sum_{k=0}^{n-1}\cos(\omega k\Delta
t)(-1)^kd \omega.
\label{deltagammacos}
\end{eqnarray}
By using the relationship \be \sum_{k=1}^n (-1)^k
\cos(kx)=-\frac{1}{2}+\frac{(-1)^n\cos\left(\frac{2n+1}{2}x\right)}
{2\cos\left(\frac{x}{2}\right)},
\label{sum_cos}\ee
we can rewrite the above equation as \ba \Delta\Gamma_n^{\text{PDD}}
&=& 2\int_0^\infty\eta(\omega)\left\{(-1)^{n+1}\cos[(n+1)\omega
n\Delta t]\right. \nonumber \\ &+& \left.(-1)^{n+1}\cos(n\omega
n\Delta t) \right. \nonumber \\ &+&
\left. 2\frac{(-1)^n\cos\left(\frac{2n+1}{2}\Delta
t\omega\right)}{\cos\left(\frac{\Delta
t\omega}{2}\right)}\right\}d\omega \nonumber \\ &=&
4\int_0^\infty\eta(\omega)\left\{(-1)^{n+1}\right. \nonumber \\
&\times& \cos\left(\frac{2n+1}{2}\Delta t\omega\right)
\left.\cos\left(\frac{\Delta t\omega}{2}\right) \right. \nonumber \\
&+& \left. \frac{(-1)^n\cos\left(\frac{2n+1}{2}\Delta
t\omega\right)}{\cos\left(\frac{\Delta
t\omega}{2}\right)}\right\}d\omega.  \ea 
\noindent 
This finally can be rearranged as
\begin{eqnarray}
\Delta\Gamma_n^{\text{PDD}}&=& 4\int^\infty_0
\eta(\omega)\sin^2\Big(\frac{\omega\Delta t}{2}\Big)\nonumber \\ &
\times &\frac{(-1)^n\cos\left[\omega\left(n+\frac{1}{2}\right)\Delta
t\right]}{\cos\left(\frac{\omega\Delta t}{2}\right)} d\omega.
\end{eqnarray}
We now take the limit of $n\to\infty$, and note that
\begin{eqnarray}
&&\lim_{n\to\infty}\frac{(-1)^n\cos\left(\omega\left(n+\frac{1}{2}\right)\Delta
 t\right)}{\cos\left(\frac{\omega\Delta
 t}{2}\right)}\frac{1}{2\pi}\nonumber\\ &=& \frac{1}{\Delta
 t}\sum_{l=0}^\infty\delta
 \left(\omega-\left(\omega_{\text{res}}+2l\omega_{\text{res}}\right)\right).
\label{sumdelta}
\end{eqnarray}
If we assume that
$\eta(\omega_{\text{res}})\gg\eta(\omega_{\text{res}}+2l\omega_{\text{res}})$
for $l>0$, which is true for sufficiently small $\Delta t$ (that is,
$\omega_{\text{res}}\gg \omega_c$), we can neglect contributions from
$l>0$ and define
\begin{eqnarray}
\Delta\Gamma_\infty^{}&=&\lim_{{ n\to\infty}}
\Delta\Gamma_n^{\text{PDD}}\label{deltagammafinal1}\\
&=&\int^\infty_08d\omega\delta(\omega-\omega_{\text{res}})
\eta(\omega)\omega_{\text{res}}\sin^2\Big(\frac{\omega\Delta
t}{2}\Big)\nonumber\\
&=&8\eta(\omega_{\text{res}})\omega_{\text{res}}. \nonumber 
\end{eqnarray}

\section{Long-time limit of $\Delta\Gamma_n^{\text{PDD}} (\tilde{t})$}
\label{derGamtildet}

By using Eq.~(\ref{Gamman-1}) and straightforward manipulations, we
can separate $\Delta\Gamma_n^{\text{PDD}} (\tilde{t})$ from
Eq.~(\ref{gammatilde}) into two parts, 
\be \Delta\Gamma_n^{\text{PDD}}
(\tilde{t})=\Delta\Gamma_n^{TI}+\Delta\Gamma_n^{\text{TD}}
(\tilde{t}),
\label{gammanPDD} 
\ee 
\noindent 
with \be\Delta\Gamma_n^{\text{TD}}(\tilde{t})=(-1)^n
\left[\Gamma_0(t_n+\tilde{t})-\Gamma_0(t_{n+1}+\tilde{t})\right],
\label{TD}
\ee and the second term \be \Delta\Gamma_n^{\text{TI}} =
2(-1)^n\bigg[\Gamma_0(t_{n+1})+2\sum_{j=1}^n
(-1)^{j}\Gamma_0(t_{n+1}-t_j)\bigg], 
\label{TI}\ee 
\noindent 
independent of $\tilde{t}$.  By using
\be\Gamma_0(t)=2\int_0^\infty\eta(\omega)[1-\cos(\omega t)]d\omega ,
\label{Ga0}\ee
\noindent 
we can rewrite $\Delta\Gamma_n^{\text{TD}} (\tilde{t})$ as
\ba\Delta\Gamma_n^{\text{TD}}(\tilde{t})&=&-4(-1)^n
\int_0^\infty\eta(\omega)\sin\Big(\frac{\Delta
t\omega}{2}\Big)\nonumber\\
&&\sin\Big\{\Big[\Big(n+\frac{1}{2}\Big)\Delta t
+\tilde{t}\Big]\omega\Big\}d\omega.
\label{TD2} \ea 
\noindent 
The last term in the integrand above is fast oscillating for large
$n$, so we will have that for $n>n_{\text{sat}}$ \be
\left|\Delta\Gamma_n^{\text{TD}}(\tilde{t})\right|<\epsilon,
\label{limitTD}\ee 
where $\epsilon$ can be made arbitrarily small.

Let us now consider $\Delta\Gamma_n^{\text{TI}}$. By using
Eq.~(\ref{Ga0}), the relation Eq.~(\ref{sum_cos}) and some tedious but
straightforward manipulations, we can rewrite Eq.~(\ref{TI}) as \ba
\Delta\Gamma_n^{\text{TI}} &=& -(-1)^n
4\int_0^\infty\eta(\omega)\cos[\omega (n+1)\Delta t]d\omega
\nonumber\\ &+& 4 \int_0^\infty\eta(\omega)
\frac{(-1)^n\cos\left(\frac{2n+1}{2}\Delta
t\omega\right)}{2\cos\left(\frac{\Delta
t\omega}{2}\right)}d\omega. 
\label{TI2} \ea 
\noindent 
Again, the integrand in the first term of the above equation is fast
oscillating for large $n$, while the second term tends to $\Delta
\Gamma_\infty$ for $n\to\infty$ (see Eq.~(\ref{sumdelta}).  We can then write
that for $n>n_{\text{sat}}$, \be |\Delta\Gamma_n^{\text{TI}}- \Delta
\Gamma_\infty|<\epsilon
\label{limitTI}.\ee

By combining Eqs.~(\ref{limitTD}) and (\ref{limitTI}), we finally
obtain that for {\it any} $\tilde{t}$ and $n>n_{\text{sat}}$, \be
\Delta\Gamma_n^{\text{PDD}} (\tilde{t})\approx \Delta
\Gamma_\infty. 
\label{gammansat}
\ee

We note that in the case of supraohmic environment, by using that
$\Gamma_0(\infty)\equiv\lim_{n\to\infty}\Gamma_0(t_{n+1})=
2\int_0^\infty\eta(\omega)d\omega$ is finite, and using
Eqs.~(\ref{TD}) and (\ref{TI2}), we can recast the conditions
Eqs.~(\ref{limitTD}) and (\ref{limitTI}) as \be |\Gamma_0
(t>t_{\text{sat}})- \Gamma_0 (\infty)| < \epsilon , 
\label{gamma0infty}\ee 
\noindent
where $t_{\text{sat}}=n_{\text{sat}}\Delta t$. This emphasizes that
condition (\ref{gammansat}) applies for times at which the {\it
natural} evolution saturates to its long-term behavior.

\section{Averaging properties of interpolated DD scheme} 

We begin by casting the QD Hamiltonian Eq. (\ref{arbhamiltonian})
(with $\alpha=1/2$) in the following form:
\begin{equation}
H_1=\sigma_z\otimes B_z + \sigma_0 \otimes B_0,
\end{equation}
where $B_z$ and $B_0$ are operators acting on the phonon bath, and $\sigma_0$
and $\sigma_z$ denote the identity and the Pauli matrix acting on the exciton
qubit, respectively. This allows us to express the evolution in the
presence of the $i$-th CP cycle by the propagator
\begin{eqnarray*}
U_i^{\text{CP}}(4\Delta t_i^{\text{CP}}) = U_f(\Delta t^{\text{CP}}_i)
X U_f(2\Delta t^{\text{CP}}_i)XU_f(\Delta t^{\text{CP}}_i),
\end{eqnarray*}
where $ U_f(t)=\exp(-t H_1)$ represents free evolution for a time
$t$. If we define
\begin{equation}
H_2 \equiv  - \sigma_z \otimes B_z + \sigma_0 \otimes B_0 = X H_1 X,  
\label{hams}
\end{equation}
we can write the entire sequence propagator as a Magnus series
expansion \cite{magnus},
\begin{equation}
U(t)=\exp\sum_{i=1}^{\infty}A_i(t),     
\label{magnus}
\end{equation}
for which, in the limit of sufficiently fast control, we can only
consider the first two lowest-order terms in $\Delta t$
\cite{khodjasteh}.  Specifically (in units where $\hbar =1$):
\begin{eqnarray}
&A_1&=-i\int^t_0 dt_1H(t_1),\\ 
&A_2&=-\frac{1}{2}\int^t_0dt_1\int^{t_1}_0dt_2[ H(t_1), H(t_2)],
\label{magnuscoef}
\end{eqnarray}
where $H(t)=U^\dagger_{\text{ctrl}}(t) H U_{\text{ctrl}}(t)$ is the
time-dependent (piece-wise constant, for instantaneous pulses)
effective Hamiltonian that describes the evolution under the control
propagator $U_{\text{ctrl}}(t)$ resulting from the applied pulses
\cite{NMR,Viola:1999:2417,khodjasteh}.

For the sequence of different CP cycles described in
Sec. \ref{Constrained}, $A_1 $ is proportional to the identity
operator, and hence does not contribute to dephasing.  This is a
simple consequence of the qubit spending equal amounts of time in each
of the computational basis states. More interestingly, we find
\begin{widetext}
\begin{eqnarray*}
A_2 &=&\int_0^{t_n^{\text{CP}}+4\Delta
t^{\text{CP}}}dt_1\int_0^{t_1}dt_2[H(t_1),H(t_2)] 
=\sum_{i=1}^n\int_{t_i}^{t_i^{\text{CP}}+4\Delta
t^{\text{CP}}}dt_1\int_0^{t_1}dt_2[H(t_1),H(t_2)]\nonumber\\
&=& \sum_{i=1}^n\Big\{ \Big( \int_{t_i}^{t_i+\Delta
t^{\text{CP}}}dt_1\int_0^{t_1}dt_2 
+\int_{t_i+\Delta t^{\text{CP}}}^{t_i+3\Delta
t^{\text{CP}}}dt_1\int_0^{t_1}dt_2 
+\int_{t_i+3\Delta t^{\text{CP}}}^{t_i+4\Delta
t^{\text{CP}}}dt_1\int_0^{t_1}dt_2 \Big) \, [H(t_1),H(t_2)] \Big
\}\nonumber\\
&=&\sum_{i=1}^n\Big\{ \int_{t_i}^{t_i+\Delta
t^{\text{CP}}}dt_1\sum_{j=1}^{i-1} 2\Delta t^{\text{CP}}_j [H_1,H_2]
+\int_{t_i+\Delta t^{\text{CP}}}^{t_i+3\Delta
t^{\text{CP}}}dt_1\bigg( \sum_{j=1}^{i-1}2\Delta
t^{\text{CP}}_j+\Delta t^{\text{CP}}_i \bigg) [H_2,H_1] \nonumber\\ &&
+\int_{t_i+3\Delta t^{\text{CP}}}^{t_i+4\Delta
t^{\text{CP}}}dt_1\sum_{j=1}^{i}2\Delta t^{\text{CP}}_j[H_1,H_2]\Big\}
\nonumber\\
&=&\sum_{i=1}^n\Big\{ \Delta t^{\text{CP}}_i\sum_{j=1}^{i-1}2\Delta
t^{\text{CP}}_j[H_1,H_2] 
+2\Delta t^{\text{CP}}_i\bigg(\sum_{j=1}^{i-1} 2\Delta
t^{\text{CP}}_j+\Delta t^{\text{CP}}_i\bigg) [H_2,H_1]
+\Delta t^{\text{CP}}_i\sum_{j=1}^{i}2\Delta
t^{\text{CP}}_j[H_1,H_2]\Big\} \nonumber\\ &=& 0,
\end{eqnarray*}
\end{widetext}
again up to irrelevant pure-bath terms.  This confirms the
second-order cancellation claimed in the main text.



\begin{thebibliography}{199}
\bibitem{remark2} An important difference between CDD protocols
constructed out of two-axis controls (as in
Ref. \onlinecite{khodjastehprl}) vs.  single-axis controls (as
examined here) stems from the fact that the relevant DD group does
not act irreducibly in the latter case. While group reducibility 
can, in general, worsen CDD performance \cite{Lea}, this is not
expected a play a role in purely dephasing settings, since all the
higher-order system-bath Magnus terms are proportional to $\sigma_z$.
\end{thebibliography}
\end{document}